\documentclass[superscriptaddress,twocolumn,secnumarabic,nofootinbib]{revtex4-1}
\usepackage{amsmath}
\usepackage[utf8]{inputenc}
\usepackage[T1]{fontenc}

\pdfoutput=1

\usepackage{bm}
\usepackage{times}
\usepackage{braket}
\usepackage{color,graphicx}
\usepackage[utf8]{inputenc}
\usepackage[T1]{fontenc}
\usepackage[export]{adjustbox}

\usepackage{enumitem}

\usepackage{xcolor}

\newcommand{\blue}[1]{\color{blue} #1 \color{black}}

\usepackage{lipsum}

\usepackage[normalem]{ulem}

\definecolor{nicered}{rgb}{0.7,0.1,0.1}
\definecolor{nicegreen}{rgb}{0.1,0.5,0.1}

\usepackage{hyperref}
\hypersetup{colorlinks,citecolor=nicegreen,linkcolor=nicered,urlcolor=black}
\hypersetup{colorlinks=true}

\begin{document}

{\flushright
{\blue{ \hfill}\\
\blue{ \hfill}\\
\blue{IFT-UAM/CSIC-19-3}\\
\blue{FTUAM-19-1}}\\
\hfill 
}

\title{Flavor constraints on electroweak ALP couplings}

\author{M.B.~Gavela} 
\author{R.~Houtz} 
\author{P.~Quilez} 
\author{R.~del Rey}

\affiliation{Departamento de Física Teórica and Instituto de Física Teórica, IFT-UAM/CSIC,
Universidad Autónoma de Madrid, Cantoblanco, 28049, Madrid, Spain}

\author{O.~Sumensari} \email[]{belen.gavela@uam.es} \email[]{rachel.houtz@uam.es}  \email[]{pablo.quilez@uam.es} \email[]{rocio.rey@uam.es}\email[]{olcyr.sumensari@pd.infn.it}
\affiliation{Dipartimento di Fisica e Astronomia ``G.\ Galilei'', Universit\` a di Padova, Italy \\
Istituto Nazionale Fisica Nucleare, Sezione di Padova, I-35131 Padova, Italy}  
  
\begin{abstract}

We explore the  signals of axion-like particles (ALPs) in flavor-changing neutral current (FCNC) processes. 
The most general effective linear Lagrangian  for ALP couplings to the electroweak bosonic sector is considered, 
and  its contribution  
to FCNC decays is computed 
	 up to
one-loop order. 
The interplay between the different couplings opens new territory for experimental exploration, as analyzed here in the ALP mass range $0<m_a \lesssim 5$ GeV. 
 When kinematically allowed,  $K\to \pi \nu \bar{\nu}$ decays provide the most stringent constraints for channels with invisible final states,  while  $B$-meson decays are  more  constraining for visible decay channels,  such as displaced vertices in $B\to K^{(\ast)}  \mu^+ \mu^-$ data. The complementarity with collider constraints is discussed as well. 

\end{abstract}
\maketitle

\section{Introduction}
\label{sec:intro
}
 With the Higgs discovery, an era in humankind's quest for the fundamental laws of Nature has been completed~\cite{Aad:2012tfa,Chatrchyan:2012xdj}. At the same
time, 
 new uncharted territory has been opened: the spin-zero window to
the universe.   
(Pseudo)Nambu-Goldstone scalars (pGBs) are strongly motivated from
fundamental problems of the known particle physics laws, that is, of the Standard Model of Particle Physics (SM). 
They are the generic tell-tale of exact, although ``hidden'' (i.e. spontaneously broken), global symmetries of nature.  A paradigmatic example is the 
 axion, which results from the dynamical solution to the strong CP problem of the SM~\cite{Peccei:1977hh,Kim:1979if,Shifman:1979if,
Zhitnitsky:1980tq,Dine:1981rt}. 
 The traditional ``invisible axion'' is expected to be extremely light, with mass $m_a<10^{-2}$ eV, and its   
scale $f_a$ to be out of direct experimental reach, although recently tantalizing axion solutions to the strong CP problem are being explored with scales as low as
 $\mathcal{O}($TeV$)$~\cite{Rubakov:1997vp,Fukuda:2015ana,Berezhiani:2000gh,Hsu:2004mf,Hook:2014cda,Chiang:2016eav,Dimopoulos:2016lvn,Gherghetta:2016fhp,Kobakhidze:2016rwh,Agrawal:2017ksf,Agrawal:2017evu,Gaillard:2018xgk}. PGBs of deep interest extend well beyond axions,  though. They appear in a plethora of 
constructions which reach beyond the SM (BSM),  typically as SM scalar singlets, e.g.~in: i) Theories with extra space-time dimensions, 
  ii)  Dynamical explanations to the smallness of neutrino masses: the Majoron~\cite{Gelmini:1980re}, iii)  String theory  models~\cite{Cicoli:2013ana}, iv) Supersymmetric extensions of the SM~\cite{Bellazzini:2017neg}, and v) Many dynamical flavor theories  with hidden global  $U(1)$ symmetries, a particular class of which  
identifies the QCD axion  as a flavon ``\emph{\`a la} Froggat-Nielsen'' (axiflavon or flaxion)~\cite{Wilczek:1982rv,Ema:2016ops,Calibbi:2016hwq}.   These pGBs  are often  denoted by the general name of axion-like
particles (ALPs), as  anomalous couplings to gauge currents often appear in addition to purely derivative ones. ALPs may or may not have anomalous couplings to gluons, and they are not required to solve the strong CP problem. One practical difference between a generic ALP and true axions which solve the strong CP problem is that, for ALPs,  
$f_a$ and  $m_a$ are treated as independent  parameters. 
 Outstandingly, and as a wonderful byproduct, both axions and ALPs  may  be  excellent candidates to explain the nature of dark matter (DM)~\cite{Abbott:1982af, Dine:1982ah,Preskill:1982cy}. 
 
The parameter space for very light ALPs, with masses below the MeV scale, is delimited mainly by astrophysical and cosmological constraints.  
 Regarding terrestrial experiments,  ADMX has finally entered the critical territory expected for an invisible axion signal if  DM is made of axions.  {In addition, the investment in axion
and ALP searches in a large range of masses is accelerating at present with CAST~\cite{Anastassopoulos:2017ftl}, IAXO~\cite{Armengaud:2014gea,Giannotti:2016drd}, and 
 future projects such as Madmax, CASPEr, QUAX, HeXenia, FUNK and electric dipole moment searches (PSI and Co)~\cite{Majorovits:2017ppy,Budker:2013hfa,Ruoso:2015ytk,Experiment:2017icw,Zenner:2013rta}. Also, DM
experiments like Xenon~\cite{Aprile:2017lqx} and the future Darwin~\cite{Aalbers:2016jon} target keV mass ALP dark matter (and solar axions). On the precision arena, flavor  experiments  provide  
 valuable
constraints. For instance, NA62~\cite{NA62:2017rwk}  is taking data, and new fixed target facilities (e.g. SHIP~\cite{Alekhin:2015byh}) are in
preparation, with sensitivity to MeV-GeVs ALPs and strong complementary potential to tackle ALP couplings to
gauge bosons and fermions. Belle-II~\cite{DePietro:2018sgg} will also have some sensitivity to this mass range, as well as the LHC
with Mathusla, Faser and CodexB~\cite{Alpigiani:2018fgd,Feng:2018noy,Gligorov:2017nwh}.
Indeed, ALPs may well show up first at colliders~\cite{Jaeckel:2015jla,Mimasu:2014nea}. Intense work on ALP signals at  the LHC and future colliders is underway~\cite{Brivio:2017ije,Bauer:2017ris}, and the synergy between collider and low-energy fixed target experiments  is increasingly explored~\cite{Izaguirre:2016dfi}.  All couplings must be analyzed  combining fixed-term and accelerator data in a complementary approach.

In this work, we explore ALP  contributions to flavor changing neutral current (FCNC) processes, formulating them in a model-independent  approach via the linear realization of the ALP effective Lagrangian.  The complete basis of  bosonic and CP-even  ALP couplings to the electroweak sector is considered. That is,  the set of  gauge invariant and independent leading-order couplings 
to the $W$, $Z$, photon and Higgs doublet is discussed. Given  that these operators  are flavor blind, they may impact flavor-changing data only at loop level.   The couplings of ALPs to  heavy SM bosons had been largely disregarded until recently, even if {\it a priori} they are all expected to be on equal footing with the pure photonic ones because of  electroweak  gauge invariance. In addition to novel  collider signatures~\cite{Brivio:2017ije,Bauer:2017ris}, rare hadron decays provide a superb  handle on the ALP couplings
to massive vector bosons~\cite{Alonso-Alvarez:2018irt} for ALP masses below 5 GeV.  
 The one-loop impact on  FCNC processes of the anomalous ALP-$W$-$W$ coupling was first considered in Ref.~\cite{Izaguirre:2016dfi}: it was shown to 
induce flavor-changing rare meson decays via $W$ exchange, with the ALP
radiated from the W boson \cite{Hiller:2004ii,Freytsis:2009ct,Izaguirre:2016dfi}.  The axion can then either decay in some visible channel or escape the detector unnoticed, 
and novel bounds  were  derived in both cases.   
   Given the level of accuracy provided by present  flavor experiments,  it is most pertinent to take into account the  
competing  contribution of other electroweak ALP couplings leading to the same final states. In other words, the ensemble of the linearly independent ALP-electroweak couplings should be considered simultaneously in order to delimitate the parameter space. Putative anomalous couplings of ALPs to gluons could also contribute to flavor-blind decays into visible channels, but not to  FCNC processes other than 
via  pseudoscalar (e.g. ALP-$\eta'$ and  ALP-pion) mixing in SM flavor-changing decays, and they are not considered in this paper. 
   
  The analysis of two (or more) couplings simultaneously has the potential to  change the experimental perspective on ALPs. Our theoretical analysis is confronted with the prospects for ALP detection in present and upcoming fixed-target experiments and $B$-physics experiments.  
After the theoretical analysis, the structure of this paper reflects successively the two alternative scenarios mentioned above, in which the ALP produced in FCNC meson decays can then either decay into visible channels within the detector, or it can be invisible by escaping the detector (or decaying to a hidden sector). 
 For both  cases, the comparison with data  considers first each coupling separately and then the ensemble in combination, and the resulting interference patterns are worked out in detail.

\section{Bosonic ALP lagrangian}
\label{sec:lagrangian}
  The most general effective Lagrangian describing ALP couplings contains -- at leading order in the linear expansion -- only three independent operators involving electroweak gauge bosons~\cite{Georgi:1986df,Choi:1986zw, Salvio:2013iaa,Brivio:2017ije},  
\begin{align}
\begin{split}
\delta \mathcal{L}_{\mathrm{eff}} &= \dfrac{\partial_\mu a\, \partial^\mu a}{2} - \dfrac{m_a^2 \,a^2}{2} \\[0.4em]
&+ {c_{a\Phi}} \,\mathcal{O}_{a\Phi}+{c_B}\, \mathcal{O}_B+{c_W} \,\mathcal{O}_W\,,
\end{split}
\label{eq:lagrangian-0}
\end{align}
\noindent with 
\begin{align}
\begin{split}
\mathcal{O}_{a\Phi} &\equiv i\dfrac{\partial^\mu a}{f_a}\, \Phi^\dagger \overleftrightarrow{D}_\mu \Phi \,,\\[0.4em]
\mathcal{O}_B&\equiv - \dfrac{a}{f_a}B_{\mu\nu} \widetilde{B}^{\mu\nu} \,,\\[0.4em]
\mathcal{O}_W &\equiv - \dfrac{a}{f_a}W_{\mu\nu}^a \widetilde{W}^{\mu\nu}_a \,,
\label{operators}
\end{split}
\end{align}
\noindent where $\Phi$ is the SM Higgs doublet, $f_a$ is the ALP decay constant, $c_i$ are real operator coefficients and $\Phi \overleftrightarrow{D}_\mu \Phi \equiv \Phi^\dagger \big{(}D_\mu\Phi\big)-\big{(}D_\mu \Phi\big{)}^\dagger \Phi {}$. The dual field strengths are defined as   $\tilde{X}^{\mu\nu}\equiv \frac12 \epsilon^{\mu\nu\rho\sigma}X_{\rho\sigma}$, with $\varepsilon^{0123}=1$. 

Upon electroweak symmetry breaking, $\mathcal{O}_{a\Phi}$ induces a mixing between $a$ and the would-be Goldstone boson eaten by the $Z$. Its physical impact is best illustrated via an ALP-dependent rotation of the Higgs field, namely $\Phi \to \Phi \, e^{i c_{a\Phi} a/f_a}$~\cite{Georgi:1986df}, which trades $\mathcal{O}_{a\Phi}$  for the following fermionic couplings: 
\begin{align}
\label{eq:leff-alp}
\mathcal{O}_{a\Phi} \rightarrow  i\, \dfrac{a}{f_a} \Big{[}\overline{Q}\, Y_u \widetilde{\Phi}\, u_R -  \overline{Q}\, Y_d \Phi\, d_R - \overline{L} \, Y_\ell \Phi \,\ell_R\Big{]}+\mathrm{h.c.}\,,
\end{align}
 where $Y_{u,d,\ell}$ denote the SM Yukawa matrices, flavor indices are omitted, and neutrino masses are disregarded. 
  The ALP-electroweak operators in Eq.~(\ref{operators}) are flavor blind, but  $\mathcal{O}_{a\Phi}$  and  $\mathcal{O}_{W}$ can participate  in FCNC processes at one loop via $W^\pm$ gauge boson exchange. 
  At this order,  the parameter space of ALP-electroweak  couplings in FCNC processes is thus reduced to 
  two dimensions  spanned by the  coefficients
  \begin{equation}
  \{c_W\,, {c_{a\Phi}}\}\,.
  \label{cparam}
  \end{equation}
 They may contribute to rare decays as illustrated in   the left (${c_{a\Phi}}$) and right (${c_W}$) panels of 
 Fig.~\ref{fig:separate-constraints}.   While ${c_W}$ has been discussed separately in Ref.~\cite{Izaguirre:2016dfi,Alonso-Alvarez:2018irt}, and  the  effective ALP-fermionic interactions have also been considered by themselves before~\cite{Hiller:2004ii,Freytsis:2009ct,Dolan:2014ska,Batell:2009jf,Bauer:2017ris,Bauer:2018uxu}, the interplay between $c_W$ and $c_{a\Phi}$ will be shown below to lead to interesting new features.

\section{FCNC ALP interactions}
\label{sec:fcnc}

The effective interaction between a pGB and left-handed fermions can be expressed in all generality as
\begin{equation}
\mathcal{L}_\text{eff}^{d_i \to d_j} = -g_{ij}^a\, \left(\partial_\mu a\right) \, \bar{d_j} \gamma^\mu P_L d_i+\mathrm{h.c.}\,, 
\label{current}
\end{equation}
\noindent 
where latin indices $i,j$ denote flavor and $g_{ij}^a$ is an effective coupling.  

The impact of  $\mathcal{O}_{a\Phi}$  and  $\mathcal{O}_{W}$ on $d_i \to d_j a$ (with $i\neq j$) transitions  via one-loop $W^\pm$ exchange induces a left-handed current of the form in Eq.~(\ref{current}), and thus a contribution to  rare meson decays.  The corresponding Feynman diagrams at the quark level are those contained in the illustration in     
 Fig.~\ref{fig:diagrams}, as well as  the corresponding self-energy diagrams with the ALP operator inserted on the 
 quark lines external to the $W$ loop. 
 At the quark level, those one-loop $W$ exchanges result in a contribution to $g_{ij}^a$, for $i\neq j$, given by
\begin{equation}
\label{eq:ga-loop}
g_{ij}^a = g^2\sum_{q=u,c,t} \dfrac{V_{qi} V_{qj}^\ast}{16\pi^2} \Bigg{[} \dfrac{3c_W}{f_a}g(x_q)-\dfrac{c_{a\Phi}}{4f_a} x_q \log \left(\dfrac{f_a^2}{m_q^2}\right)\Bigg{]}\,,
\end{equation}
where $g$ is the electroweak gauge coupling, and $V_{qi}$  are the CKM  matrix elements. In this equation, $m_q$ denotes the mass of a given up-type quark $q$ that runs in the loop,  
the approximation $m_{d_j} ,\, m_{d_i} \ll m_W$  has been used, $x_q=m_q^2/m_W^2$, and the loop function is given by
\begin{equation}
g(x)=\dfrac{x\left[1+x (\log x-1)\right]}{(1-x)^2}\,.
\end{equation}
\begin{figure}[t!]
  \centering
    \includegraphics[width=0.25\textwidth]{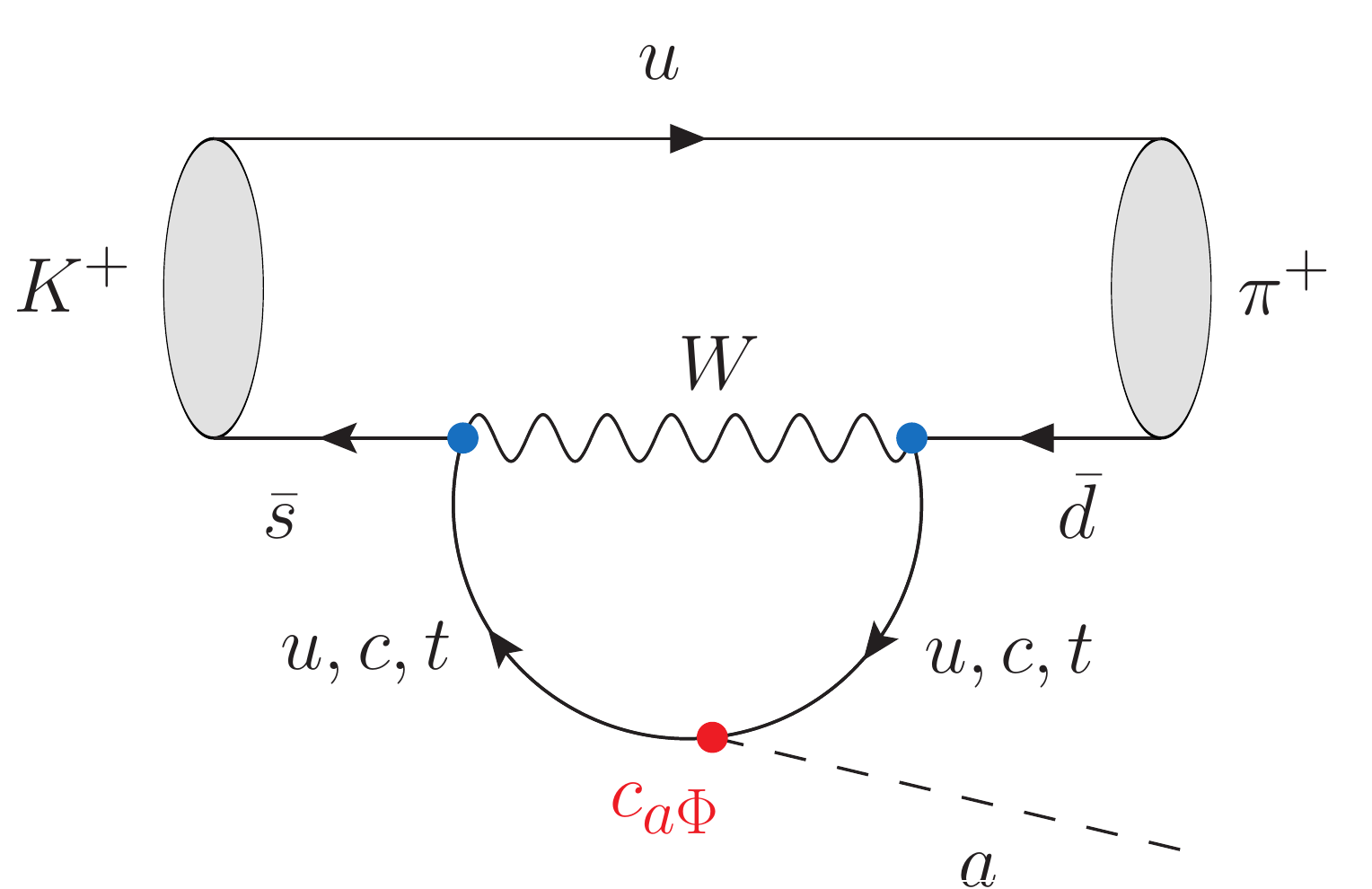}~\includegraphics[width=0.25\textwidth]{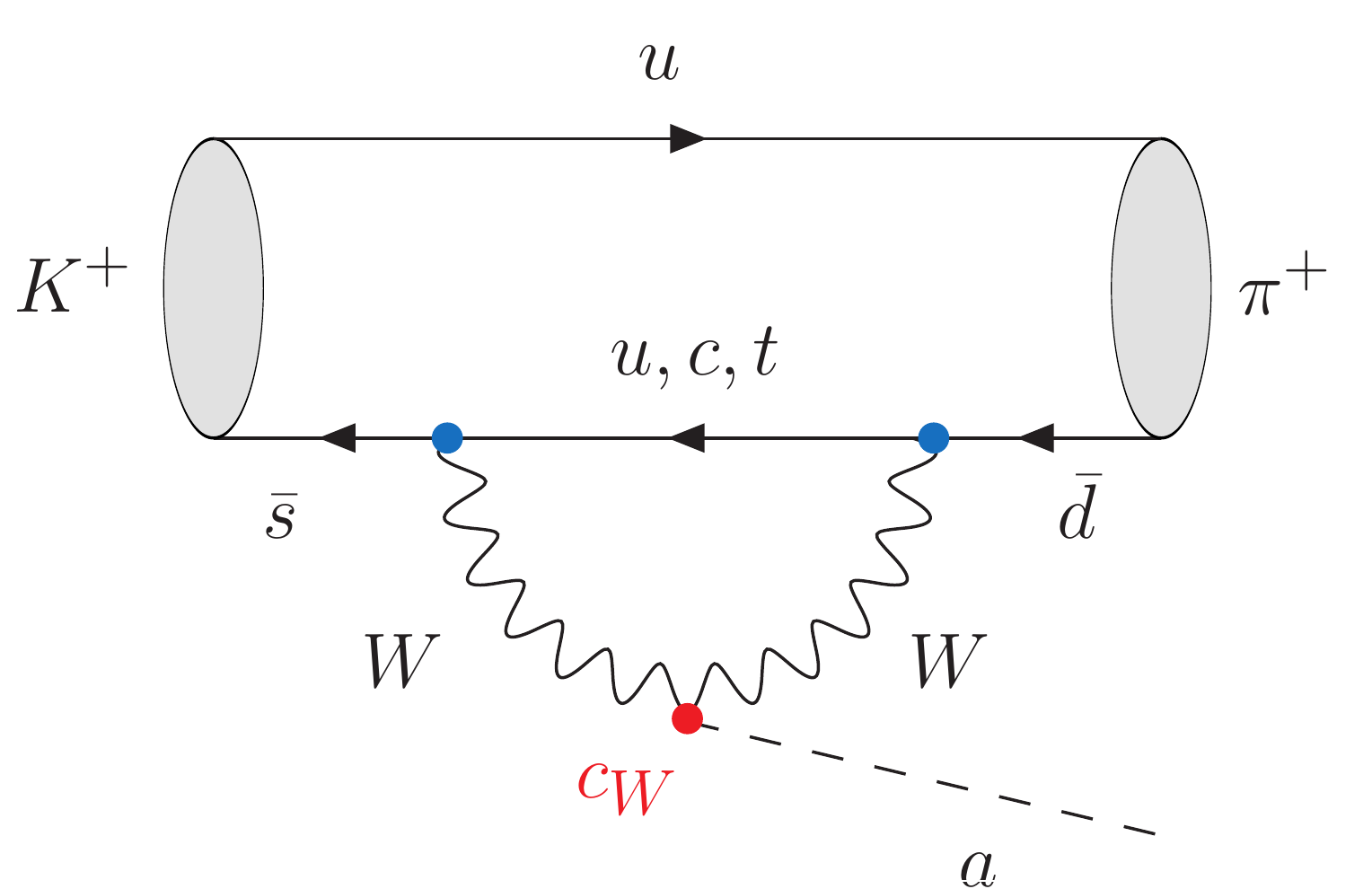}
  \caption{ \sl \small Illustration of diagrams giving one-loop contributions to the process $K^+\to \pi^+ a$ via the interactions defined in Eq.~\eqref{eq:leff-alp}. }
  \label{fig:diagrams}
\end{figure}

\begin{figure*}[t]
    \includegraphics[width=0.495\textwidth]{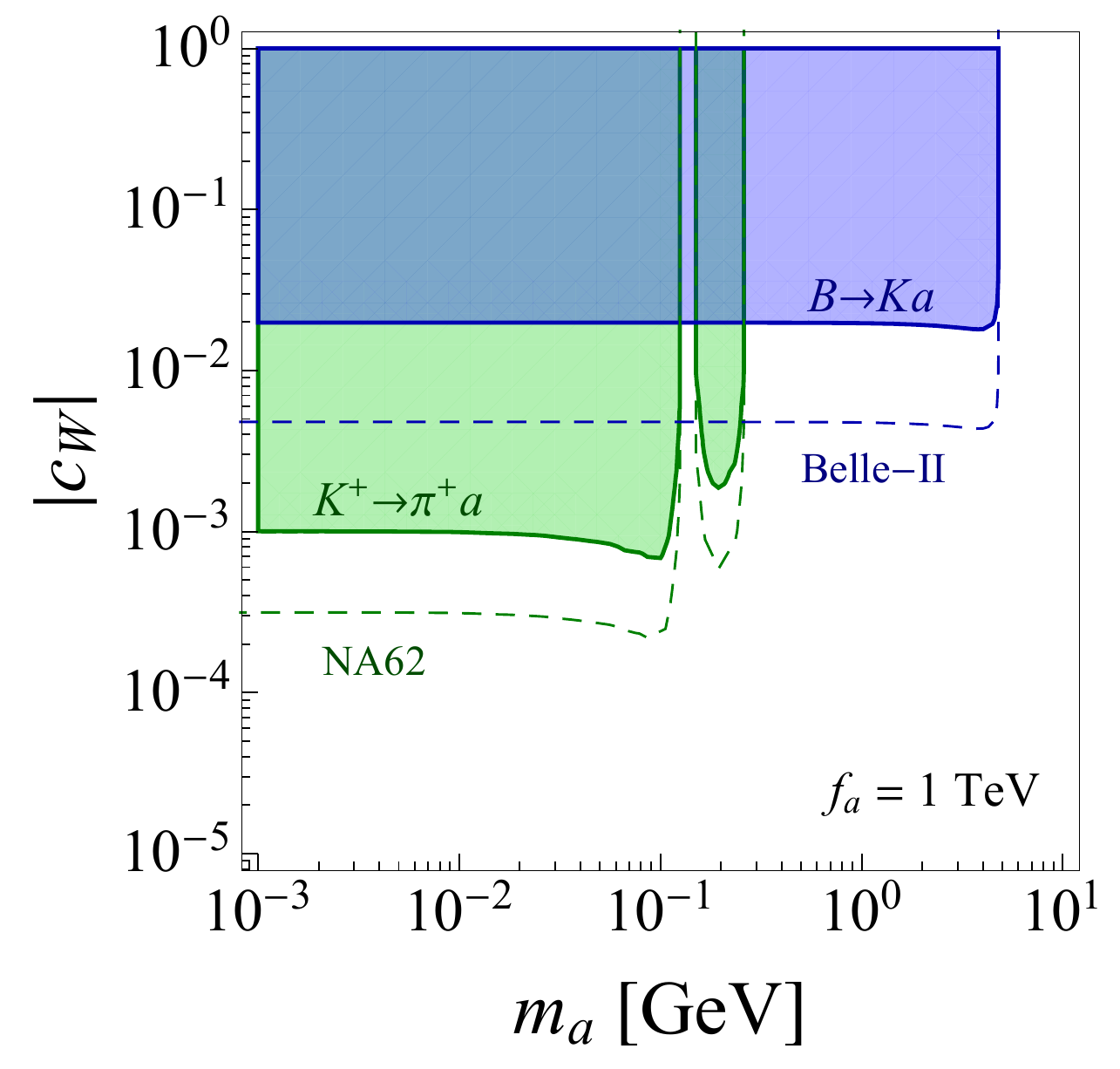}  \centering
    \includegraphics[width=0.495\textwidth]{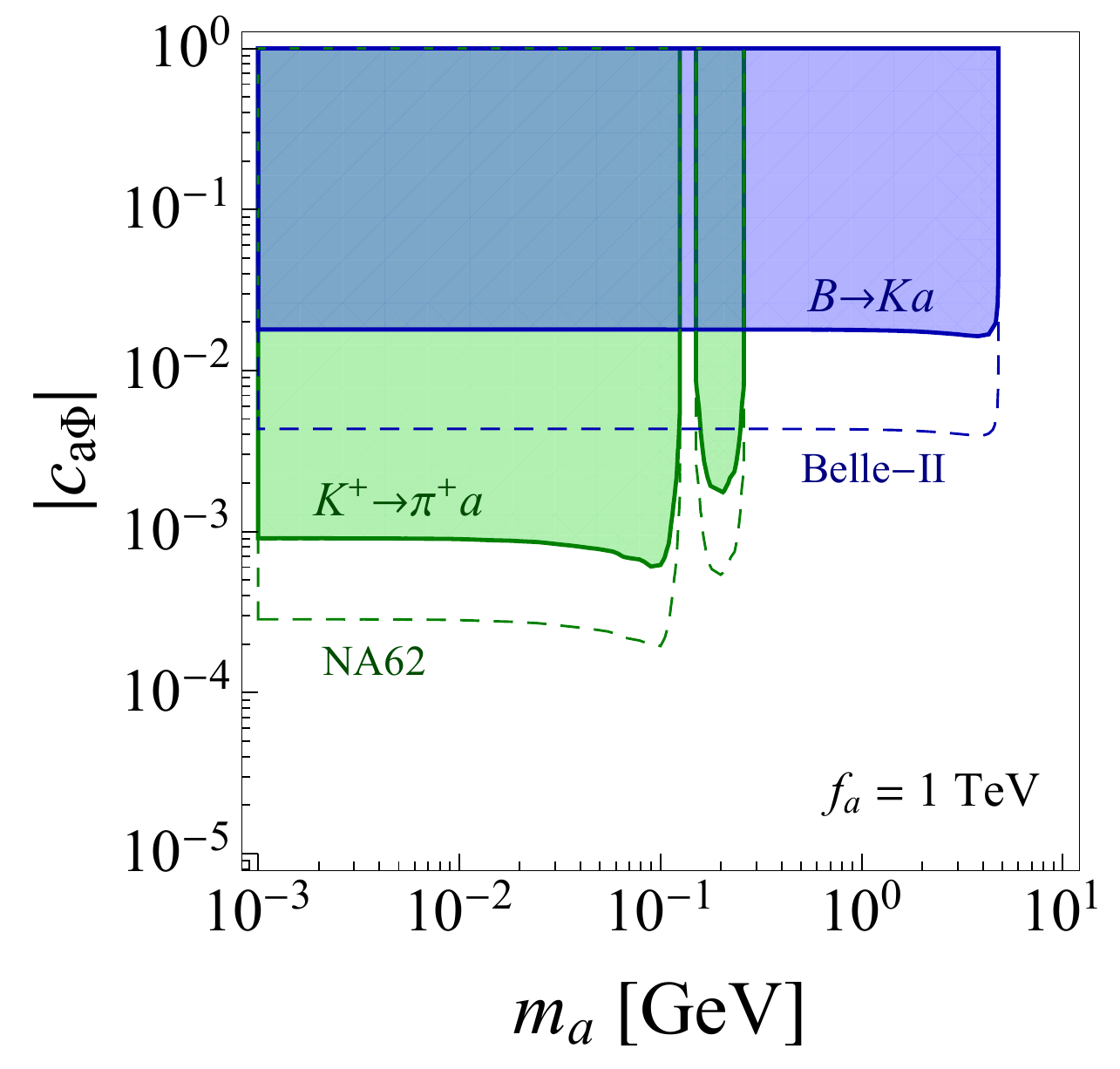}  \centering
  \caption{ \sl \small Invisible ALP: constraints on  the absolute value of $c_W$ (left panel) and $c_{a\Phi}$ (right panel) as a function of the ALP mass, considering each of these couplings separately. The exclusion contours have been derived from the experimental limits on $\mathcal{B}(K^+\to \pi^+ +\mathrm{inv})$~\cite{Artamonov:2008qb} (green) and $\mathcal{B}(B\to K +\mathrm{inv})$~\cite{Adler:2004hp}  (blue) by fixing $f_a = 1$~TeV and by setting the other couplings to zero. Projections for NA62~\cite{Izaguirre:2016dfi} and Belle-II~\cite{Kou:2018nap} experiments are illustrated by dashed lines.}
  \label{fig:separate-constraints}
\end{figure*}
It follows that the decay rate for the process $K^+\to \pi^+ a$  can be expressed  as 
\begin{align*}
\label{eq:BKpia}
\Gamma(K^+\to \pi^+ a) &=\dfrac{m_{K}^3 |g_{sd}^a|^2}{64\pi}  f_0(m_a^2)^2 \lambda_{\pi a}^{1/2}\left(1-\dfrac{m_{\pi}^2}{m_{K}^2}\right)^2\,,
\end{align*}

\noindent with $\lambda_{\pi a}=\left[1-\frac{{( m_a}+m_\pi)^2}{m_K^2}\right]\left[1-\frac{({m_a}-m_\pi)^2}{{m_K}^2}\right]$. In this expression, $f_0$ denotes the $K\to \pi$ scalar form factor, which has been computed in lattice QCD  in Ref.~\cite{Carrasco:2016kpy}. An analogous expression can be obtained \textit{mutatis mutandis} for the decay $B\to K a$, in which case the relevant form factors can be found in Refs.~\cite{Ball:2004ye,Straub:2015ica}.

In Eq.~(\ref{eq:ga-loop}), the contribution  proportional to $c_W$ is finite due to the Glashow–Iliopoulos–Maiani  (GIM) mechanism, in agreement with the results of Ref.~\cite{Izaguirre:2016dfi}. The $c_{a\Phi}$  term is instead logarithmically sensitive to the ultraviolet scale of the theory $f_a$,  
and its contribution is thus approximated by  the leading log   
model-independent component.
 Furthermore,  because $g(x)\sim x + \mathcal{O}(x^2)$ for small $x$, the contributions from the up and charm quarks are sub-leading  in both terms   with respect to that of the top quark. 
Also, note that the   logarithmic enhancement of the  $c_{a\Phi}$ term  ($\propto \log \left(f_a/m_t\right)$) should be particularly relevant for large values of $f_a$. This logarithmic divergence is a consequence of  the operator $\mathcal{O}_{a\Phi}$ being non-renormalizable~\cite{Freytsis:2009ct,Batell:2009jf,Dolan:2014ska}, in contrast with renormalizable scenarios such as two-Higgs doublet models \cite{Frere:1981cc,Freytsis:2009ct,Li:2014fea,Arnan:2017lxi}.

The interplay between $c_{a\Phi}$ and $c_W$ presents interesting features which depend on their relative sign.  Their  
contributions to ALP production  in rare decays can interfere destructively if and only if $c_{a\Phi}/c_{W}>0$. Such a cancellation would leave a region in parameter space which cannot be probed by relying only on FCNC decays such as $K\to \pi a$ and $B\to K a$. An alternative to lift this degeneracy using LHC constraints will be discussed further below, after deriving the constraints that follow from rare meson decays.

 In order to determine the detection possibilities for a given final state channel, an important
element is whether the ALP can decay into visible particles within the detector, or  whether it escapes and contributes to
an ``invisible'' channel. We discuss next both cases.

\section{The invisible ALP}
\label{sec:invisible}
Let us consider first the scenario of an ALP that does not decay into visible particles in the detector, which we shall refer to as the ``invisible ALP''. 
This situation can arise if $a$ is sufficiently light, making $a$ long-lived, or if there are large couplings of $a$ to a dark sector, making $\mathcal{B}(a\to \mathrm{inv})$ sufficiently large. 
 The analysis performed below is general and applies to both cases.

\begin{figure*}[t]
    \includegraphics[width=0.495\textwidth]{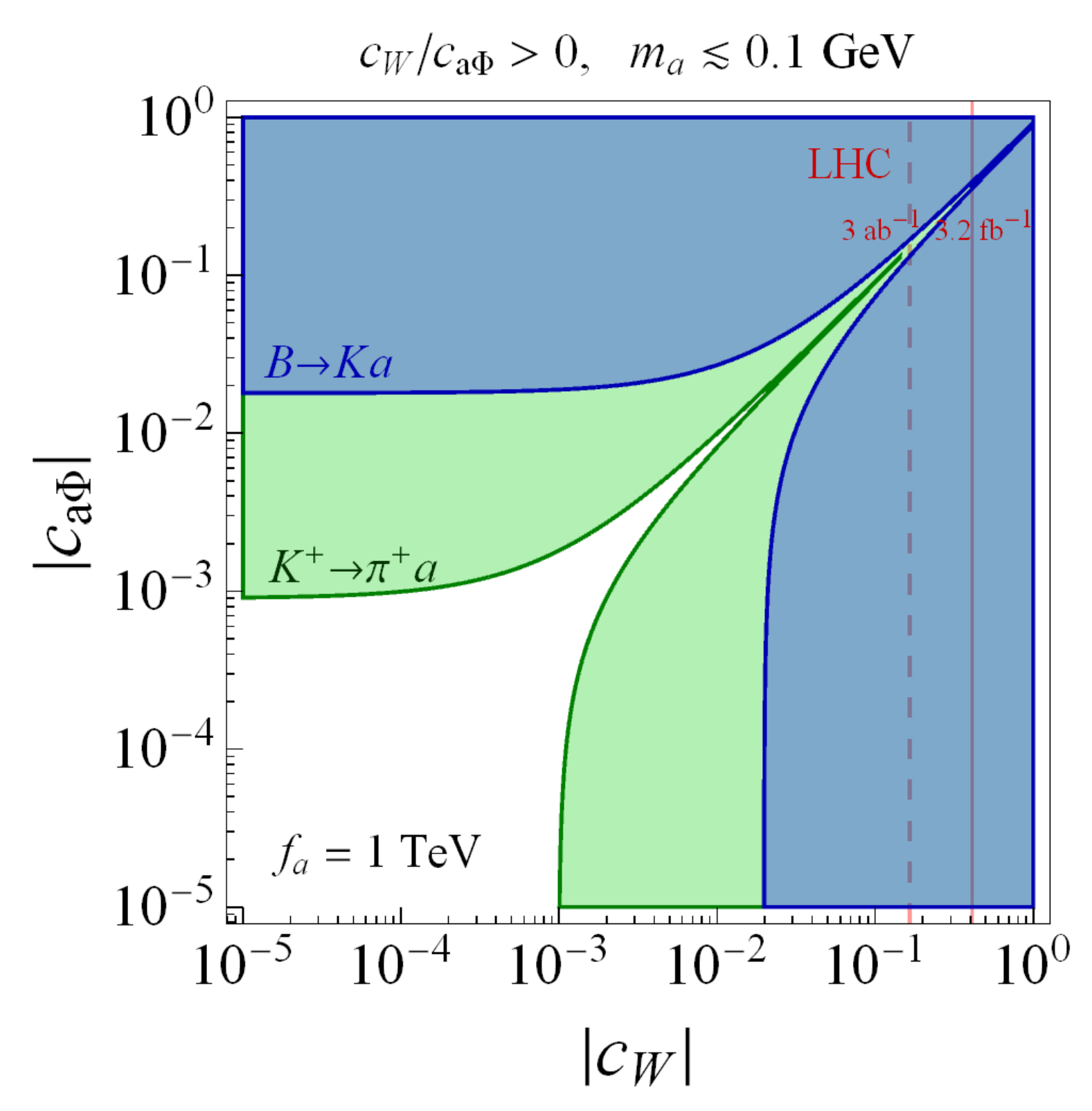}  \centering 
    \includegraphics[width=0.495\textwidth]{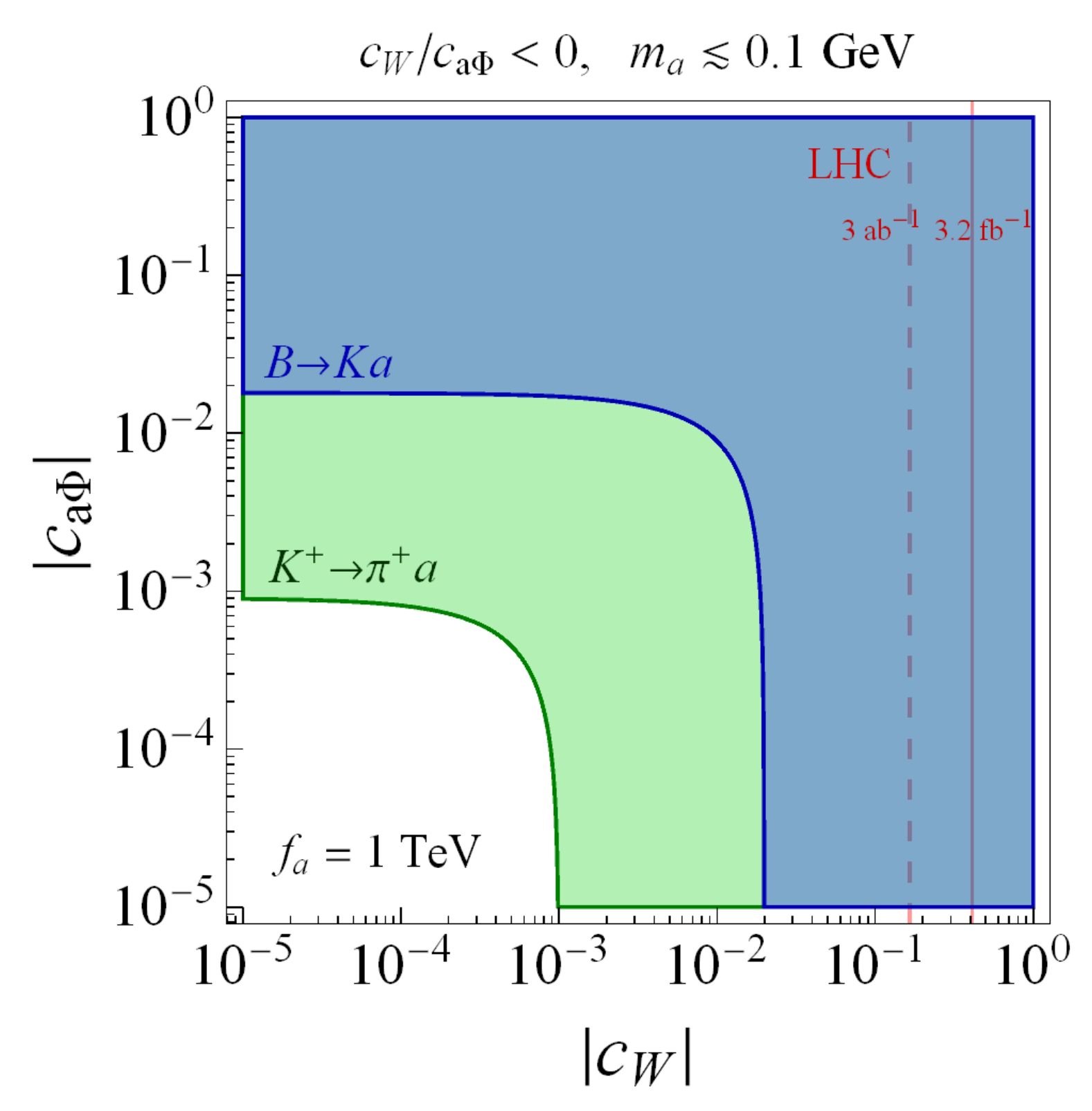}  \centering   
  \caption{ \sl \small  Allowed $\{c_W, c_{a\Phi}\}$ parameter space for the invisible ALP when those two couplings are simultaneously present.   The superposition of the constraints from  $K^+\to \pi^+ + \mathrm{inv} $ (green) and $B^+\to K^+ + \mathrm{inv} $ (blue) data is shown for an illustrative case with $f_a=1$~TeV and $m_a \lesssim 100$~MeV. The left (right) panel shows the destructive (constructive) interference of the two couplings for $c_{W}/c_{a\Phi}>0$ ($c_{W}/c_{a\Phi}<0$). The red solid (dashed) lines correspond to the current (projected) limits from mono-$W$ searches at the LHC with $3.2~\text{\sl \small fb}^{-1}$ ($\,3~\text{\sl \small ab}^{-1}\!$)  of data~\cite{Brivio:2017ije}. 
   }
  \label{fig:combined-constraints}
\end{figure*}

The experimental constraints relevant for different $m_a$  ranges are listed next:

\begin{itemize}[leftmargin=*]
\item[•] $m_a \in (0, m_K-m_\pi)$: \\[0.3em]
Searches for the decay $K\to \pi \nu\bar{\nu}$ have been performed at the E787 and E949 experiments. The bounds obtained can be directly reinterpreted to limit the parameter space of new undetected particles.  
E787 and E949 experiments take measurements in two regions of pion momentum, namely $p_\pi\in (140,199)$~MeV and $p_\pi\in (211,229)$~MeV, 
 which can be translated into the ALP mass ranges $150~\mathrm{MeV}\lesssim m_a \lesssim 260$~MeV and $m_a \lesssim115$~MeV, respectively. The limits reported in these searches are $\mathcal{B}(K^+ \to \pi^+ \nu \bar{\nu})^{\mathrm{exp}}=\left(1.73^{+1.15}_{-1.05}\right) \times 10^{-10}$~\cite{Artamonov:2008qb} and $\mathcal{B}(K^+ \to \pi^+ \nu \bar{\nu})^{\mathrm{exp}}< 2.2 \times 10^{-9}$ \cite{Adler:2004hp}, which lie slightly above the SM prediction, $\mathcal{B}(K^+\to \pi^+\nu\bar{\nu})^{SM}= \left(9.11\pm 0.72\right) \times 10^{-11}$~\cite{Buras:2015qea}. Similar searches have been performed at the NA62 experiment, which aims at attaining the SM rates in the very near future~\cite{CortinaGil:2018fkc}. In our analysis, we consider the E787 and E949 constraints, as summarized in Ref.~\cite{Artamonov:2008qb}.
\item[•] $m_a \in (0, m_B-m_K)$: \\[0.3em]
The most constraining experimental limits on $\mathcal{B}(B\to K^{(\ast)}+\mathrm{inv})$ were obtained by the Belle collaboration. These are $\mathcal{B}(B\to K \nu \bar{\nu})<1.6 \times 10^{-5}$ and $\mathcal{B}(B\to K^\ast \nu \bar{\nu})<2.7 \times 10^{-5}$ (90$\%$ C.L.)~\cite{Grygier:2017tzo}, which lie respectively a factor of 3.9 and 2.7 above the SM predictions~\cite{Buras:2014fpa}. In the near future, Belle-II aims at measuring the SM value with a $\mathcal{O}(10\%)$ precision~\cite{Kou:2018nap}. For the new physics scenario  considered here, the strongest constraint arises from the $B\to K \nu\bar{\nu}$ result.
\end{itemize}
We have explicitly checked that $\Delta F=2$ constraints on the effective couplings $g_{ij}^a$ are less stringent than the ones presented above for most of the ALP parameter space considered here. Nevertheless, 
those constraints should provide the best bounds on $c_{a\Phi}$ and $c_{W}$ for masses larger than $\sim 5$ GeV, which are out of reach of rare decays, see Fig.~\ref{fig:separate-constraints}. Those observables  are not included in our analysis, though, since the consistent assessment of the corresponding limits would require a complete two-loop computation, as well as the additional consideration of higher dimension ALP operators, which goes beyond the scope of this paper. 


The constraints set on ALP-electroweak coefficients by data  will be analyzed in two steps: first within  a {\it one coupling at a time} approach, where  either only $c_W$ {\it or} 
   $c_{a\Phi}$  are switched on; next, the $\{c_{a\Phi}, c_W \}$ parameter space spanned by the {\it simultaneous presence of both couplings} will be considered. 
   
   Fig.~\ref{fig:separate-constraints} depicts the allowed values of $c_W$ (left panel) and $c_{a\Phi}$ (right panel) as a function of the ALP mass, when only one of these two couplings is added to the SM. 
    The constraints obtained on the $\{ m_a, c_W\}$ plane (left panel)  coincide with those derived in Ref.~\cite{Izaguirre:2016dfi}. 
     The constraints on the parameter space for $\{ m_a, c_{a\Phi}\}$ (right panel) are a novel contribution of this work. The case illustrated corresponds to $f_a=1$~TeV. The quantitative similarity of the exclusion limits on the two couplings depicted in Fig.~\ref{fig:separate-constraints} is fortuitous; it is easy to check that  the constraints on $c_{a\Phi}$ become stronger than those for $c_W$ for larger values of  $f_a$, as expected from the logarithmic dependence of its contribution, see Eq.~(\ref{eq:ga-loop}).
 
    These plots  also indicate that  kaon constraints are typically one order of magnitude stronger than those derived from $B$-meson decays, although limited to a more restricted $m_a$ range.  Future prospects from NA62 and Belle-II  are also illustrated in Fig.~\ref{fig:separate-constraints}  with dashed lines.

When both $c_{a\Phi}$ and $c_W$ are simultaneously considered, an interesting pattern of destructive interference 
 can take place, as anticipated in Sec.~\ref{sec:fcnc}.  
  Fig.~\ref{fig:combined-constraints} depicts the result of combining the different experimental constraints for fixed values of $f_a$ and $m_a\lesssim 0.1$~GeV.  This  shows indeed that 
 when the relative sign of both couplings is positive, a blind direction in parameter space appears.  
  This unconstrained direction  is exactly  aligned for kaon and $B$-meson decays. For this reason, additional experimental information is then needed to lift the degeneracy. One possibility  is to consider the decays $D\to \pi (a\to \mathrm{inv})$, which are sensitive to a different combination of $c_{a\Phi}$ and $c_W$, since 
the up- and down-type quark contributions to the term proportional to $c_{a\Phi}$ have opposite signs, see Eq.~\ref{eq:leff-alp}.  These decays, however, suffer from a heavy GIM suppression, and no such experimental searches have been performed to our knowledge. A more promising possibility is to consider LHC constraints that are sensitive to a specific ALP coupling. For example, LHC searches for mono-$W$ final states are only sensitive to $c_W$\footnote{Bounds stemming from mono-Z signals are slightly better, but
this final state can also be generated by another
coupling ($c_B$), which complicates slightly
the reinterpretation in terms of $c_W$ and
$c_{a\Phi}$.}. In Ref.~\cite{Brivio:2017ije} the authors derived the  current (projected) bounds
\begin{equation}
\dfrac{|c_W|}{f_a}\lesssim 0.41~(0.16)~\mathrm{TeV}^{-1}\,,
\end{equation} 
from $3.2~\mathrm{fb}^{-1}$ ($3~\mathrm{ab}^{-1}$) of LHC data: these   have been superimposed in Fig.~\ref{fig:combined-constraints}. 
 Similarly, a reinterpretation of $pp \to t \bar{t}+\mathrm{MET}$ at the LHC would constrain only
$c_{a\Phi}$, but such analysis goes beyond the scope of this letter. Typically, LHC constraints are weaker than flavor bounds, except in the region of parameter space where the flavor signal is suppressed due to a cancellation between two contributions. 
In this case, the complementarity of low and high-energy constraints becomes an important handle on new physics. 
\section{The visible ALP }
\label{sec:visible}
We analyze next the case of ALPs  produced at  loop level 
via rare meson decays, but 
decaying into visible states via the same set of bosonic interactions  
 introduced in Eq.~(\ref{operators}). For the $m_a$ range considered in this work, the kinematically accessible decays are $a\to \gamma\gamma$, $a\to \mathrm{hadrons}$ and $a\to \ell \ell$, with $\ell=e,\mu,\tau$.    Both tree-level and loop-level contributions to the decays are to be taken into account.  
 Indeed,  experimental limits on ALP couplings to photons, electrons, and
nucleons are so stringent  that (indirect) loop-induced observables  can give stronger constraints than (direct) tree-level 
ones~\cite{Bauer:2017ris,Alonso-Alvarez:2018irt}. 

At  tree level, $c_W$ and $c_{a\Phi}$ contribute respectively to ALP  decays into photons and into fermions. Nevertheless, the coupling $c_B$ may also enter the game for these decays: at  tree level for the photonic channel and at  loop level for the fermionic channel. That is, 
while the parameter space for the production of an ALP via rare meson  decays  is still the two-dimensional one in Eq.~(\ref{cparam}), the whole set of  ALP electroweak couplings $\{c_{a\Phi}, c_W, c_B\}$ is relevant for the analysis of visible decay channels. For consistency, all one-loop contributions induced by these three couplings are to be taken into account. 

For instance, the partial width for ALP decay into leptons, including one-loop corrections, reads  
\begin{equation}
\Gamma (a\to \ell^+\ell^-) =  
|c_{\ell\ell}|^2\dfrac{ m_a m_\ell^2 }{8\pi f_a^2} \sqrt{1-\dfrac{4 m_\ell^2}{m_a^2}}
\label{leptchannel}
\end{equation}
where $\alpha_\mathrm{em}$ is the  fine structure constant  and $c_{\ell\ell}$ is given at one-loop order by
\begin{align}
\begin{split}
c_{\ell\ell} = c_{a\Phi}&+\frac{3\,\alpha_\mathrm{em}}{4\pi} \left(\frac{3\,c_W }{s_w^2}+ \frac{5\,c_B}{c_w^2 } \right) \log \dfrac{f_a}{m_W}\\
&+ \dfrac{6\, \alpha_{\mathrm{em}}}{\pi}\left(c_B \, c_w^2+c_W\,  s_w^2\right) \log \dfrac{m_W}{m_\ell} \,,
\end{split}
\end{align}
where  $s_w=\sin \theta_w$, $c_w=\cos \theta_w$ and $\theta_W$ denotes the weak mixing angle.  For the $a\to \gamma\gamma$  decay, the partial width reads 
 \begin{align}
\label{eq:GammaAphoton}
\Gamma (a\to \gamma\gamma)& =  |c_{a\gamma\gamma}|^2 \dfrac{m_a^3 }{4\pi f_a^2}\,,
\end{align}
 where the $c_{a\gamma\gamma}$ coupling is defined at  tree level, as
\begin{equation}
c_{a\gamma\gamma}\Big{|}_{\mathrm{tree}}\equiv c_B \,c_w^2 +c_W\,s_w^2 \,.
\label{photoncoup}
\end{equation}
Furthermore, bosonic loops give corrections to $c_{a\gamma\gamma}$ proportional to $c_W$. Fermionic loops may also induce nonzero values of $c_{a\gamma\gamma}$ at the scale $\mu = f_a$, even if the ALP has no tree-level couplings to gauge bosons, i.e.~$c_W = c_B = 0$~\cite{Bauer:2017ris}. To sum up,   both $c_W$ and $c_{a\Phi}$ induce one-loop corrections to the photonic width.  Specifically, for  $m_a\ll \Lambda_{\text{QCD}}$,  
\begin{align}
\begin{split}
c_{a\gamma \gamma }\Big{|}_\mathrm{1-loop} =  &c_W\,\Big[s_w^2\,+\frac{2\,\alpha_{\text{em}}}{\pi} B_2(\tau_W)\Big] +c_B\,c_w^2 \\
&- c_{a\Phi} \,\frac{\alpha_\mathrm{em}}{4\pi}\,\bigg( B_0 
+\frac{m_a^2}{m_{\pi}^2-m_a^2}\bigg)\,, 
\end{split}
\label{Eq:cgammaEff}
\end{align}
where $B_0$ and $B_2(\tau_f)$ are loop functions, which are detailed in Appendix~\ref{App:loopFactors}}. For $m_a\gg \Lambda_{\text{QCD}}$, the second term in the last  line  of  the above equation is absent,  since it stems from $\pi$-$a$ mixing which becomes negligible in this mass range.  
 
\begin{figure*}[t]
\includegraphics[width=0.495\textwidth]{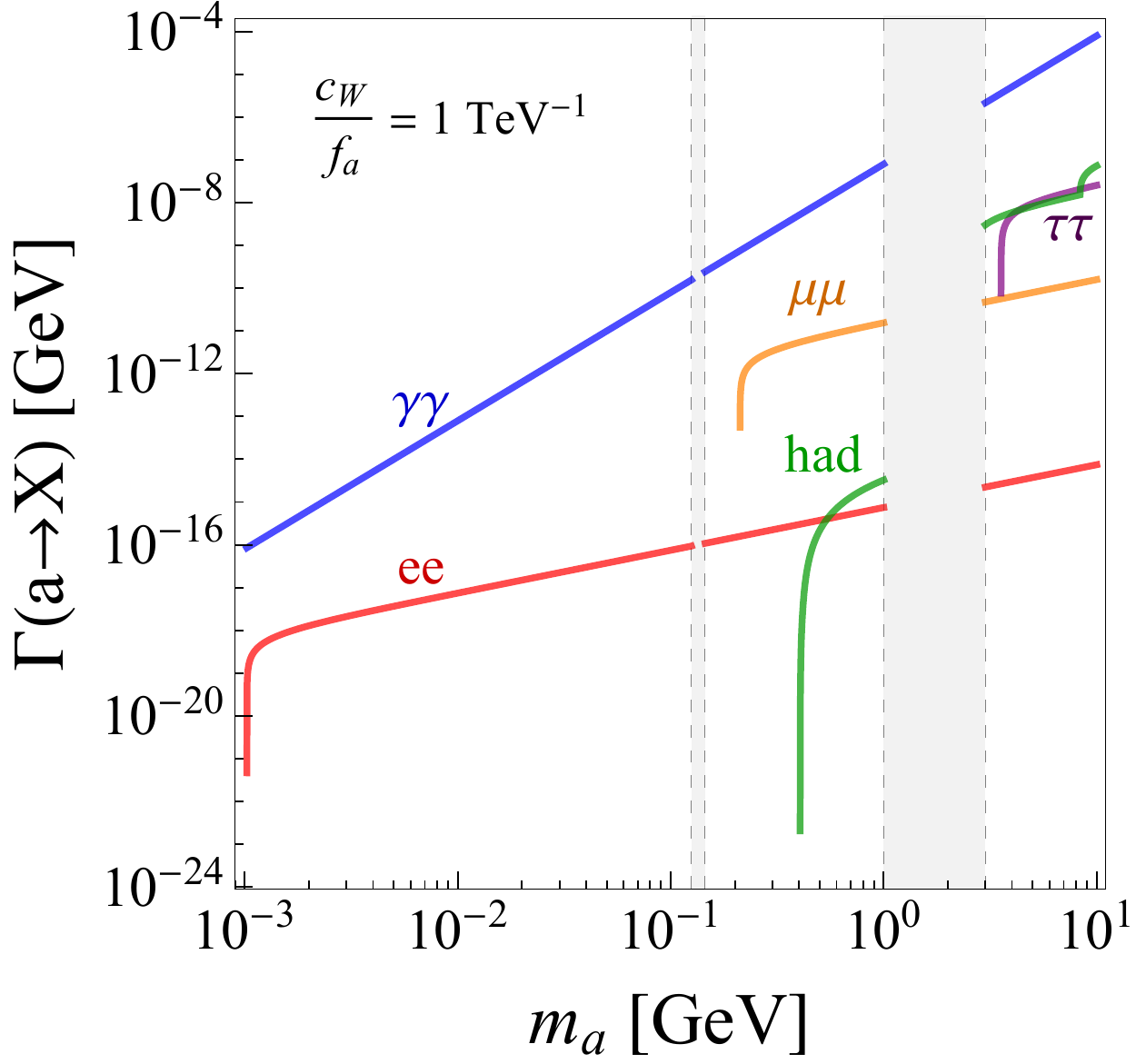} \centering 
\includegraphics[width=0.495\textwidth]{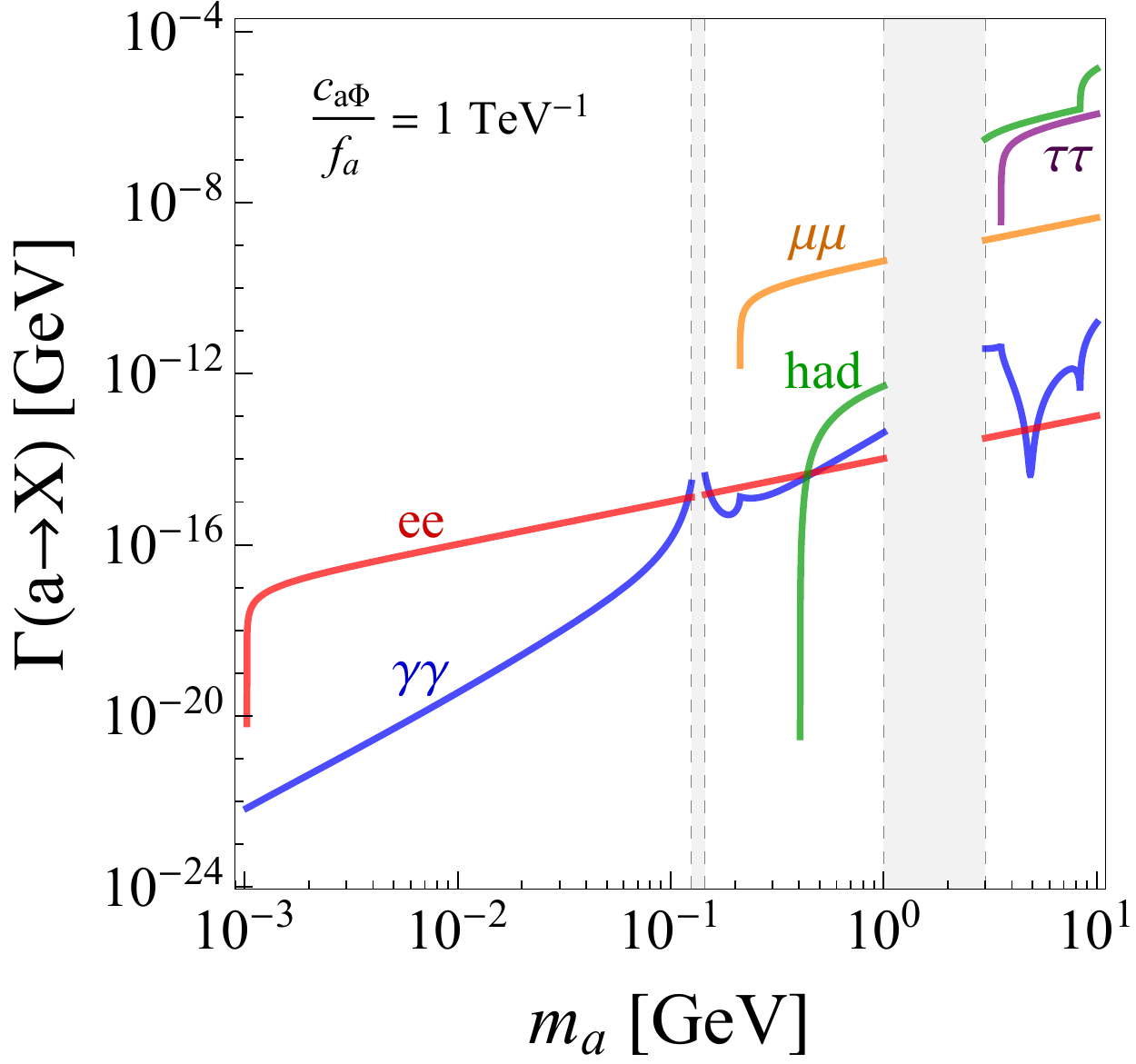} \centering 
  \caption{ \sl \small  ALP partial decay widths to various two-particle channels as a function of $m_a$, in presence of either $c_W$ (left panel) or $c_{a\Phi}$ (right panel),  for   $c_{W}/f_a=1~\mathrm{TeV}^{-1}$ and $c_{a\Phi}/f_a=1~\mathrm{TeV}^{-1}\!$, respectively.  The grey shaded areas correspond: i) to the pion mass region, which is experimentally excluded due to the large $\pi^0$-$a$ mixing; ii) the interval $(1,3)$~GeV, in which the hadronic width cannot be fully assessed either with chiral estimates or perturbatively. } 
  \label{fig:BRs}
\end{figure*}

  For hadronic decays, it is pertinent to consider two separate $m_a$ regions: (i) between $ 3\, m_\pi$ and $1$~GeV, and (ii) above $3$~GeV. In the former 
  region, the dominant hadronic decay is $a\to 3\pi$ which can be computed by employing chiral pertubation theory~\cite{Bauer:2017ris}. In the region above $3$~GeV, the dominant decays are $a\to c \bar{c}$ and $a\to b \bar{b}$, which are well described  by a perturbative expression analogous to  Eq.~(\ref{leptchannel}) multiplied by the color factor $N_c=3$.\footnote{Note that the decay $a\to gg$ is not induced at one-loop level in our setup, since the up- and down-type quark contributions cancel due to the different signs in Eq.~\eqref{eq:leff-alp}.} In this work we remain agnostic about the intermediate region $m_a\in (1,3)$~GeV, since several hadronic channels, which are particularly difficult to  estimate reliably,  open up for these masses.\footnote{A first attempt to compute these rates by using a data-driven approach in this particular $m_a$ interval has been proposed in Ref.~\cite{Aloni:2018vki} for the $G \widetilde{G}\,a$ couplings.}  In  this region,  the total hadronic width $\Gamma_a$ will be replaced by its value  at the range frontier at $m_a=3$~GeV. Note that this is the most conservative choice, since the hadronic width is a continuous and strictly increasing function of $m_a$.
  
  Fig.~\ref{fig:BRs} illustrates the ALP partial widths as a function of $m_a$, when  either only $c_W$ (left panel) or  $c_{a\Phi}$ (right panel) are present, for the benchmark values  $c_{W}/f_a=1~\mathrm{TeV}^{-1}$ and $c_{a\Phi}/f_a=1~\mathrm{TeV}^{-1}$.  The mass thresholds  for each of the fermionic channels  are clearly delineated.

 In order to analyze  the impact of an intermediate on-shell ALP on rare meson decays to visible channels, ALP production via the couplings in Eq.~(\ref{operators}) needs to be convoluted with ALP decay into SM particles via that same set of couplings.  When  $c_{a\Phi}$ and $c_W$ are simultaneously present, a very interesting pattern of constructive/destructive interference is expected. We will assume for simplicity $c_B=c_W$ to illustrate the effect. While a positive sign for
 $c_W/ c_{a\Phi}$  leads  to destructive  interference in ALP production (see Eq.~(\ref{eq:ga-loop}) and Fig.~\ref{fig:combined-constraints}),   the opposite can occur in the subsequent ALP decay  into visible channels.  Indeed,  the  decay into leptons   shows destructive interference for negative 
  $c_{a\Phi}/c_{W}$, see Eq.~(\ref{leptchannel}). The expectation for the photonic channel  is more involved and depends on the ALP mass:  for $m_a< m_{\pi}$ the terms in the last parenthesis in Eq.~(\ref{Eq:cgammaEff}) are both real and positive and the interference pattern is thus analogous to that for ALP production,  while for larger masses it may differ.  Table~\ref{tab:DestructiveVSconst} summarizes the interference pattern expected. 
 
 \begin{table}[h!]
  \renewcommand{\arraystretch}{1.7}
\begin{align*}
\begin{array}{|c|ccc|}
\hline
c_W/c_{a\phi}  & \text{Production} & a\to \ell^+\ell^-& a\to \gamma\gamma\\
\hline\hline
>0  & \text{ Destructive}    & \text{Constructive} & \text{Destructive } \\
<0 & \text{Constructive}   & \text{Destructive} & \text{Constructive}\\ \hline
\end{array}
\end{align*} 
\caption{\sl \small ALP-mediated rare meson decays:  interference pattern between $c_{a\Phi}$ and $c_W$ in ALP production and decay as a function of $c_W/c_{a\phi}$ sign, by assuming $c_B=c_W$. The $a\to\gamma\gamma$ column assumes $m_a<m_\pi$, see text for details. } 
\label{tab:DestructiveVSconst}
\end{table}

 \begin{figure*}[t]
\includegraphics[width=0.495\textwidth]{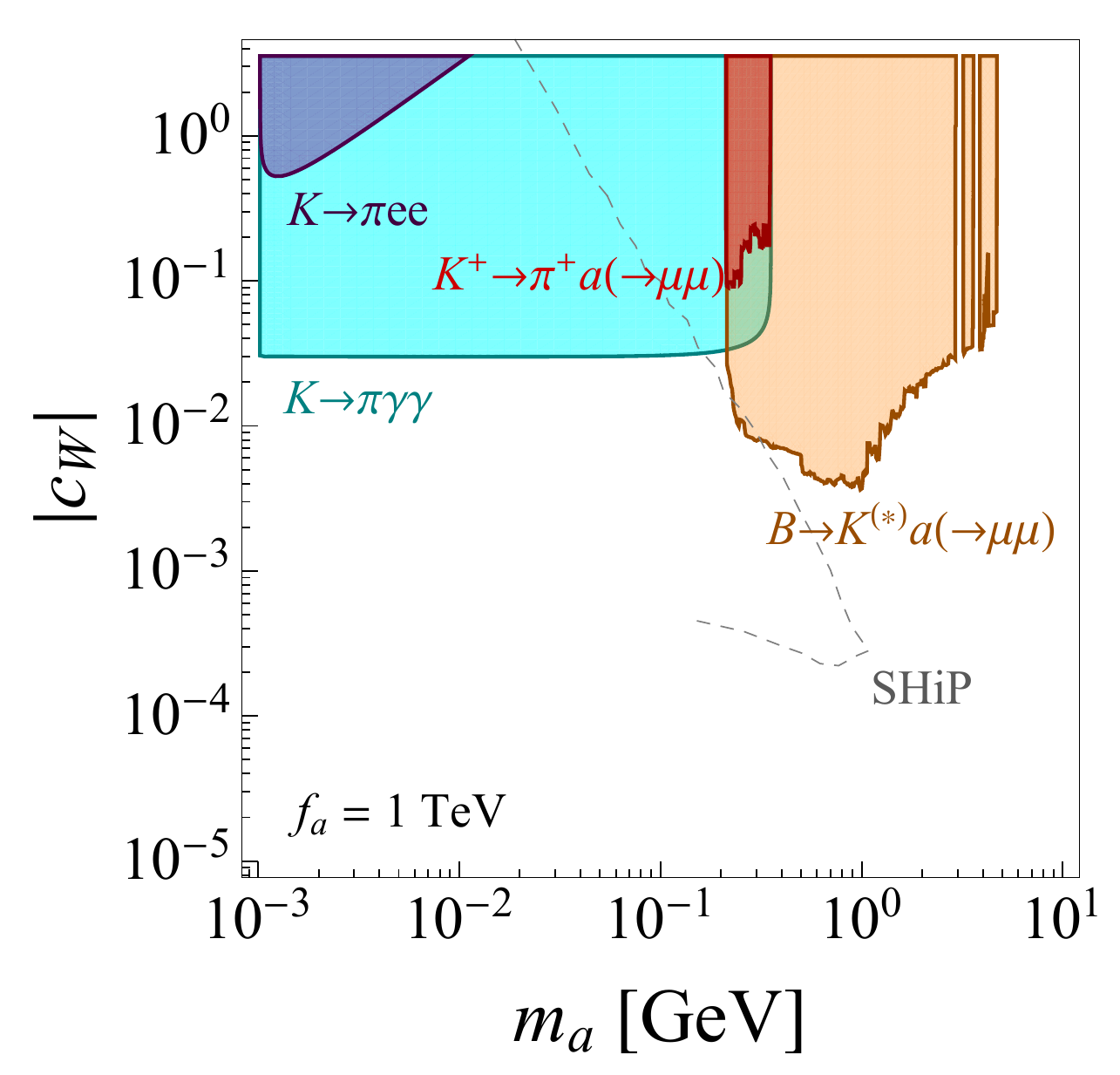}\centering 
\includegraphics[width=0.495\textwidth]{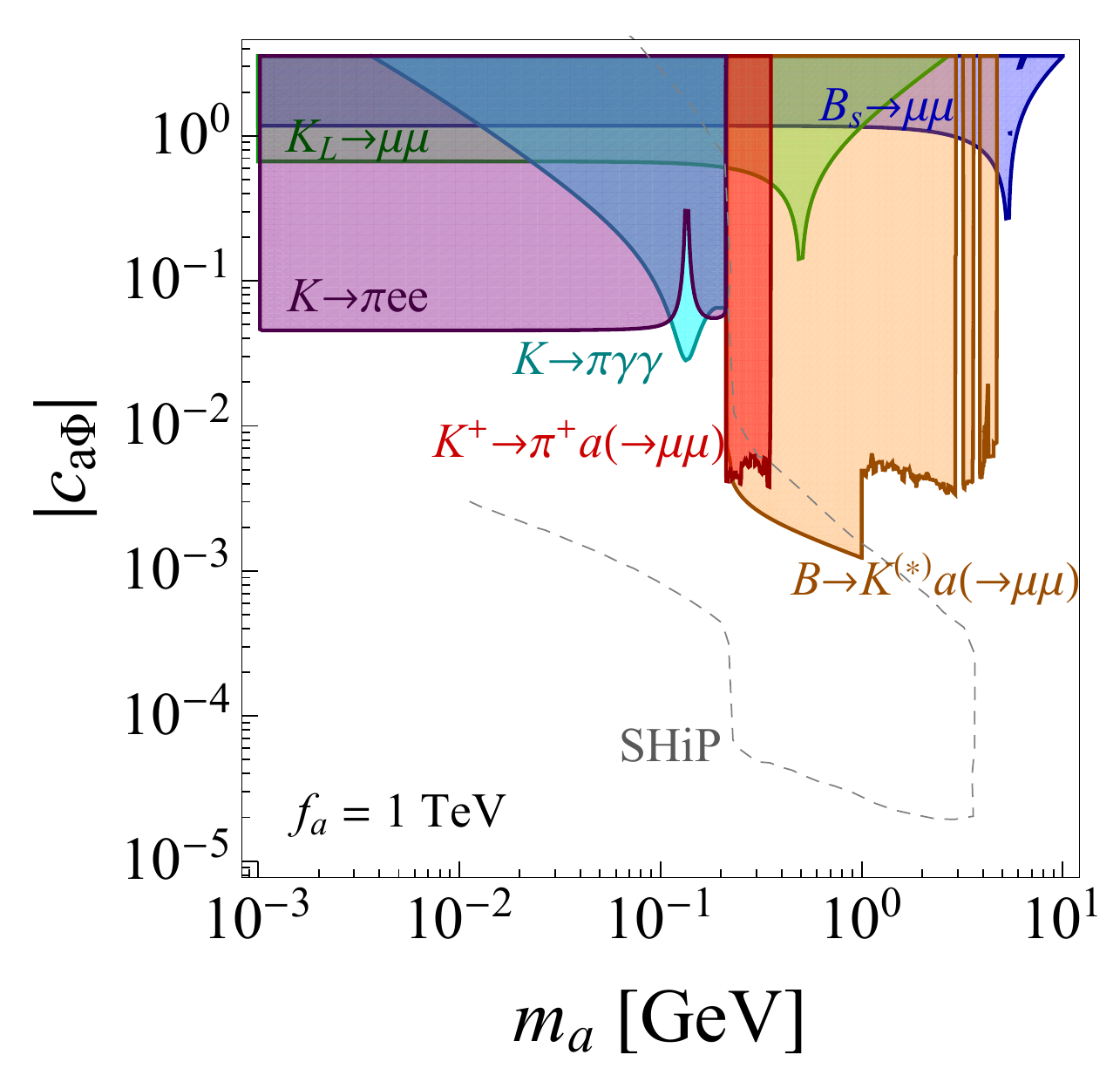}\centering 
  \caption{ \sl \small    Visible ALP: constraints on  the absolute value of $c_W$ (left panel) and $c_{a\Phi}$ (right panel) when these couplings are considered separately, 
  as a function of the ALP mass and for $f_a=1$~TeV.  The exclusion contours follow from the experimental limits on
 $K^+\to \pi^{+} \, a(\to\mu\mu)$ (red)~\cite{CERNNA48/2:2016tdo}, $B\to K^{(\ast)} \, a(\to\mu\mu)$ (orange)~\cite{Aaij:2015tna,Aaij:2016qsm}, $\mathcal{B}(K_L\to\mu\mu)$ (green)~\cite{Tanabashi:2018oca}, $\mathcal{B}(B_s\to\mu\mu)$ (blue)~\cite{Aaij:2017vad,Bobeth:2013uxa}, $\mathcal{B}(K\to \pi ee)$ (purple)~\cite{Tanabashi:2018oca} and $\mathcal{B}(K\to \pi \gamma\gamma)$ (cyan)~\cite{Artamonov:2005ru}.  The grey dashed lines are projections for the SHiP experiment~\cite{Alekhin:2015byh}.  The unconstrained regions in the range of the LHCb bounds correspond to the masses of several hadronic resonances which are vetoed in their analysis.}
  \label{fig:constraints}
\end{figure*}

 Three sets of experimental data that will be considered in order to constrain the $\{m_a,c_{a\Phi},c_{W}\}$ parameter space for a visible ALP: 1) displaced vertices; 2) semileptonic and photonic meson decays; 3) leptonic meson decays.
 \begin{enumerate}[leftmargin=*]
 \item \emph{Displaced vertices}. Of particular interest are searches for long-lived scalars, which would result in displaced vertices. Two $m_a$ ranges are pertinent:     \begin{enumerate}[leftmargin=*]
    \item $m_a \in (2 m_\mu, m_B-m_K)$ \\[0.3em]
    The LHCb collaboration perfomed searches for long-lived (pseudo)scalar particles in the decays $B\to K^{(\ast)}a$, with $a\to \mu\mu$~\cite{Aaij:2015tna,Aaij:2016qsm}. Limits on $\mathcal{B}(B\to K^{(\ast)}a)\cdot \mathcal{B}(a\to\mu\mu)$ which vary between $10^{-10}$ and $10^{-7}$ are reported  as a function of  $m_a$ and the proper lifetime, $\tau_a$. For $\tau_a<1$~ps, the limit derived  is independent of $\tau_a$ since the ALP would decay promptly. The best constraints are those for values of $\tau_a$ between $1$~ps and $100$~ps, for which  the dimuon vertex would be displaced from the interaction vertex. See also Ref.~\cite{Dobrich:2018jyi} for a recent reinterpretation of these limits.
    \item $m_a \in (2 m_\mu, m_K-m_\pi)$ \\[0.3em] Similar searches have also been performed by the NA48/2 Collaboration for the decay $K^+\to \pi^+ a$, followed by $a\to\mu\mu$ \cite{CERNNA48/2:2016tdo}. The  limits reported on $\mathcal{B}(K^+\to \pi^+ a)\cdot \mathcal{B}(a\to\mu\mu)$ decrease with  ALP lifetime until $\tau_a = 10$~ps, becoming constant for smaller values of $\tau_a$. The best experimental limits are  $\mathcal{O}(10^{-10})$ and obtained for $\tau_a \leq 10$~ps. 
    \end{enumerate}
    \item \emph{Semileptonic and photonic meson decays}. Relevant constraints on ALPs can be inferred  from their indirect contributions to  low-energy meson decays. In particular:
    \begin{enumerate}[leftmargin=*]
    \item  \emph{Kaon decays}. The measured kaon branching fractions $\mathcal{B}(K^+\to\pi^+ee)^{\mathrm{exp}}=(3.00\pm 0.09)\times 10^{-7}$, $\mathcal{B}(K^+\to\pi^+\mu\mu)^{\mathrm{exp}}=(9.4\pm 0.6)\times 10^{-8}$~\cite{Tanabashi:2018oca}, and $\mathcal{B}(K^+\to \pi^+\gamma\gamma)^{\mathrm{exp}}=(1.01\pm 0.06)\times 10^{-7}$~\cite{Artamonov:2005ru} will be taken into account.  In order to avoid the uncertainty related to the unknown SM long-distance contributions,  it will be required  that the ALP contribution alone does not saturate the $2\sigma$ experimental bounds.
    \item  \emph{B-meson decays}. Recently, LHCb observed several deviations from the expected values in ratios of $B\to K^{(\ast)}\mu\mu$ and $B\to K^{(\ast)}ee$ decays in different bins of dilepton squared mass~\cite{Aaij:2014ora,Aaij:2017vbb}. If these anomalies turn out to imply new physics,  ALP couplings would not  explain them. More precisely, pseudoscalar effective operators induced by a heavy mediator cannot reproduce current deviations due to the constraints derived from $\mathcal{B}(B_s \to \mu^+\mu^-)^{\mathrm{exp}}$~\cite{Aaij:2017vad}. On the other hand, a light ALP with $m_a \lesssim m_B-m_K$  would face stringent limits from LHCb searches for long-lived (pseudo)scalar particles in $B\to K^{(\ast)}a(\to\mu\mu)$, as mentioned above~\cite{Aaij:2015tna,Aaij:2016qsm}. For these reasons, we  leave out of our analysis the constraints   that would stem from the comparison of  exclusive $B\to K^{(\ast)}\mu\mu$ measurements with the SM expectation until further clarification is provided by the $B$-physics experiments.
    \end{enumerate}
    \begin{figure*}[t]
\includegraphics[width=0.495\textwidth]{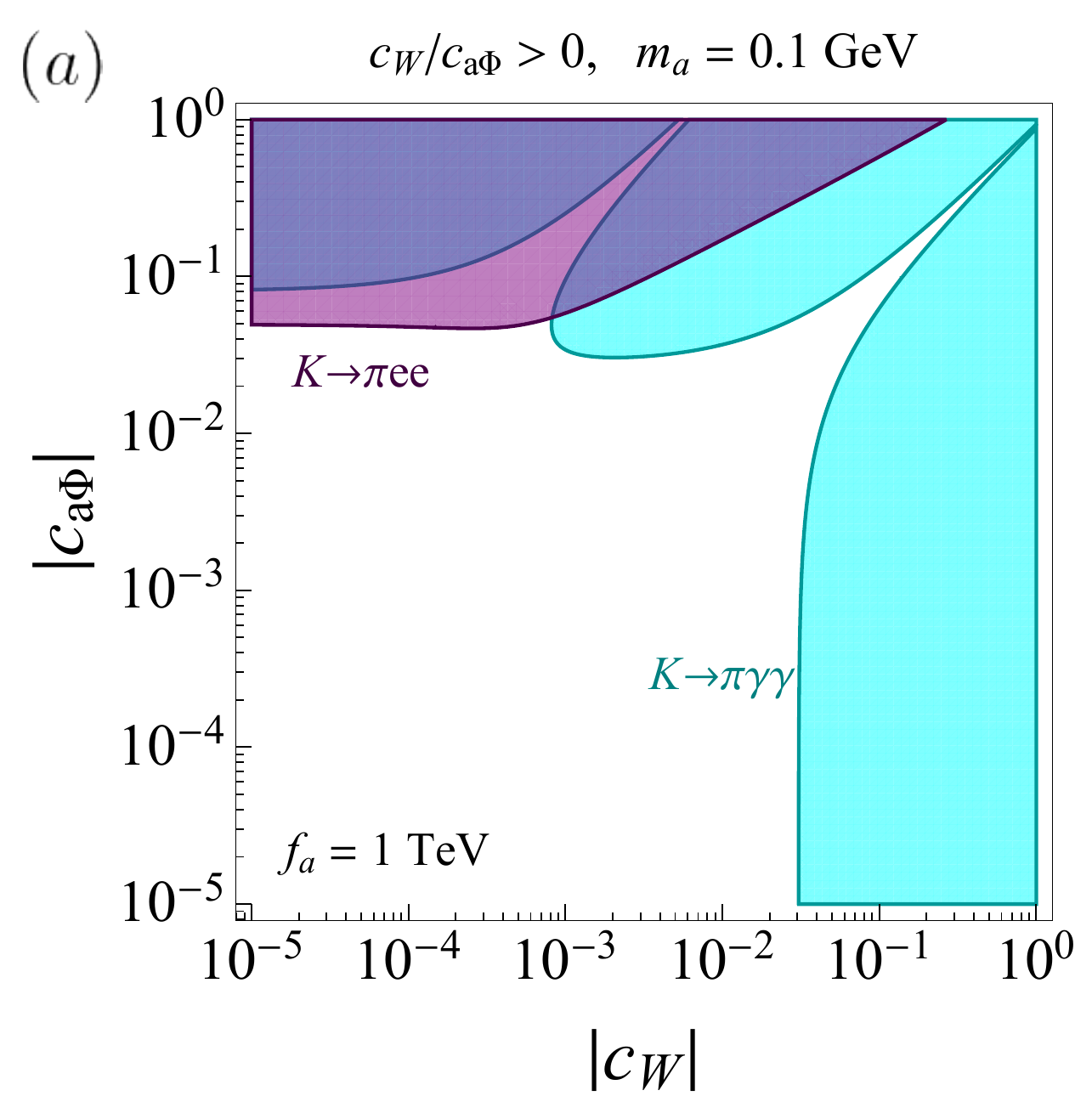}\centering
\includegraphics[width=0.495\textwidth]{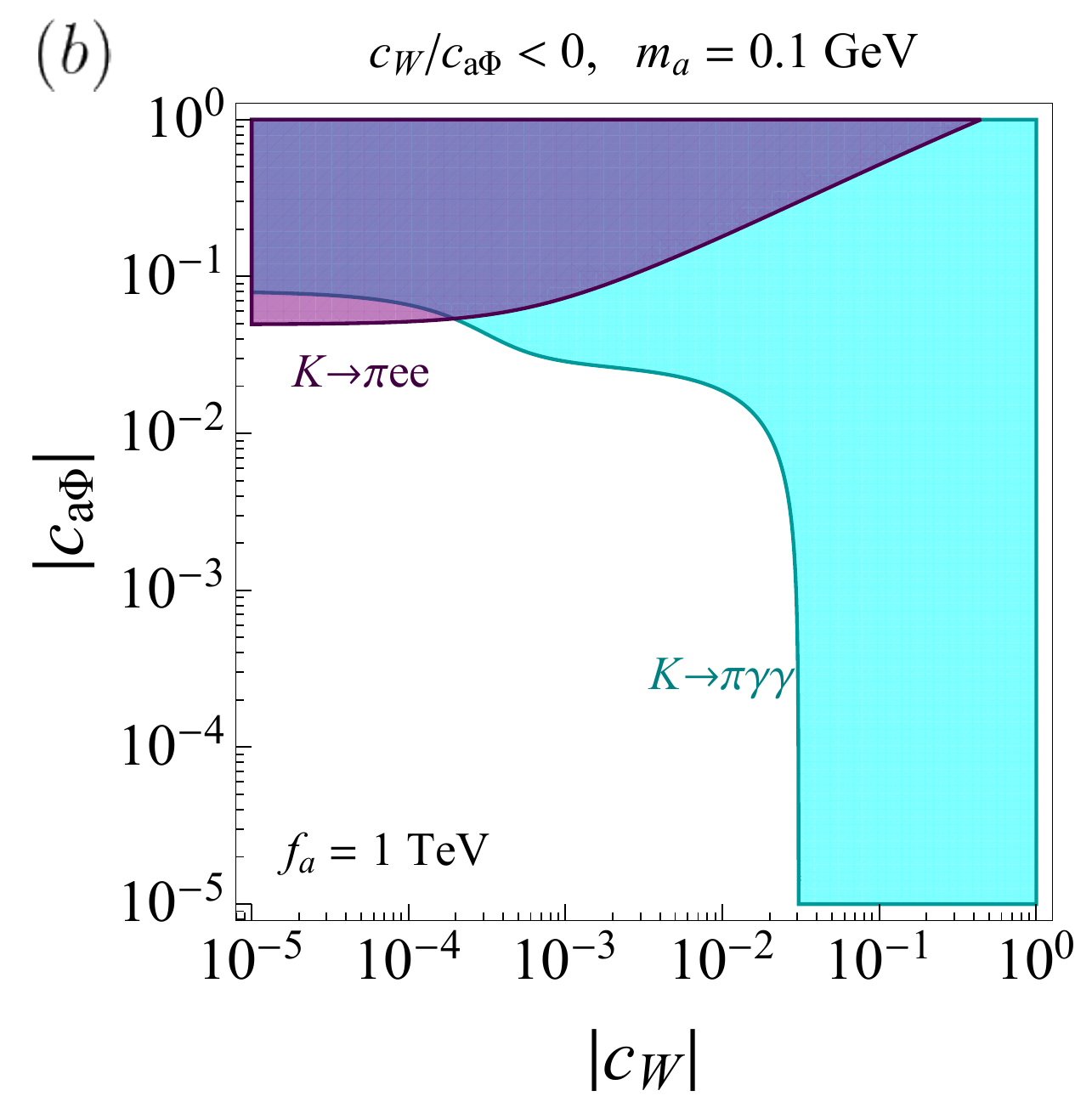}\centering\\
\includegraphics[width=0.495\textwidth]{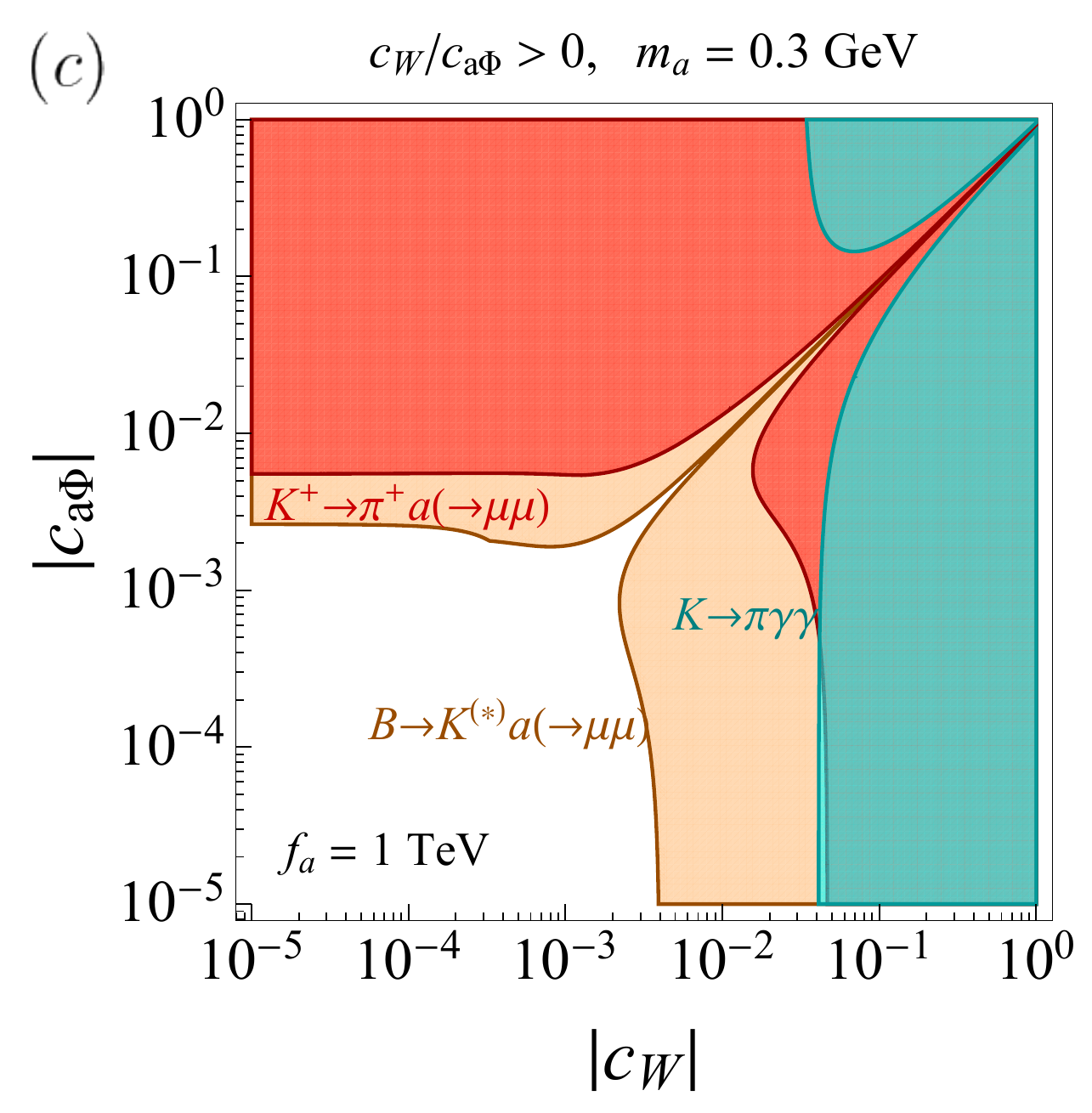}\centering
\includegraphics[width=0.495\textwidth]{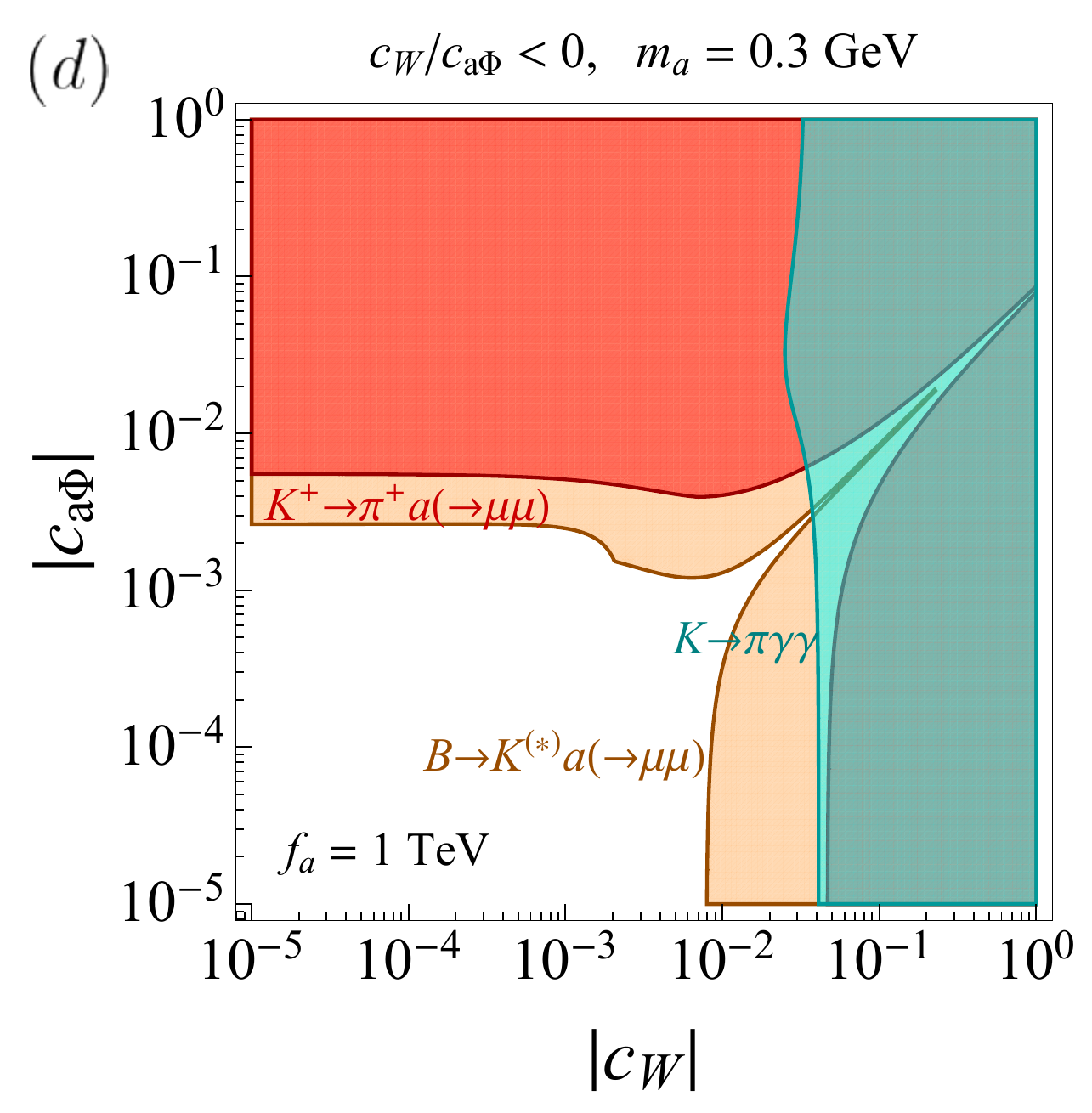}\centering 
  \caption{ \sl \small
 Visible ALP: Allowed  parameter space when the couplings $\{c_{aW},c_{a\Phi}\}$ are simultaneously present, for    $f_a=1$~TeV and $m_a=0.1$~GeV (upper plots) or   $m_a=0.3$~GeV (lower plots). The different flat directions observed in the figures correspond to the destructive interferences of both couplings in ALP production and/or  the various ALP channel decays, which depend on the sign of $c_W/c_{a\Phi}$.  See text for details. }
  \label{todovisible}
\end{figure*}
    \item \emph{Leptonic $B_s$ and $K_L$ decays}:\\[0.3em]
    While the constraints in 1) and 2) above correspond to on-shell ALPs, off-shell contributions are relevant in leptonic meson decays. 
LHCb measured $\mathcal{B}(B_s\to \mu\mu)^{\mathrm{exp}}=(3.0 \pm 0.6^{+0.3}_{-0.2})\times 10^{-9}$~\cite{Aaij:2017vad}, which agrees with the SM prediction, $\mathcal{B}(B_s\to \mu\mu)^{\mathrm{SM}}=(3.65\pm 0.23)\times 10^{-9}$~\cite{Bobeth:2013uxa}. The ALP contribution to this observable can be computed by a straightforward modification of the expressions provided in Ref.~\cite{Arnan:2017lxi}. Similarly, we consider the kaon decay $\mathcal{B}(K_L\to \mu\mu)^{\mathrm{exp}}=(6.84 \pm 0.11)\times 10^{-9}$~\cite{Tanabashi:2018oca}. In the latter case,  we impose  once again the conservative requirement that the ALP (short-distance) contribution does not saturate the $2\sigma$ experimental values. When the complete set of  electroweak couplings in Eq.~(\ref{operators}) will be  simultaneously considered for an off-shell ALP,  the interference pattern in the amplitudes can be understood  analogously to the separate discussion on production and decay  for on-shell ALPs.
 \end{enumerate}


 In analogy with the case of the invisible ALP in the previous section,  all of these data will be analyzed first within  a {\it one coupling at a time} approach,  where either only $c_W$ {\it or}  $c_{a\Phi}$  are switched on  (as $c_B$ by itself cannot mediate FCNC processes). In a second step, the {\it simultaneous presence of $\{c_{a\Phi}, c_W , c_B\}$} will be taken into account.  We assume $c_B=c_W$  in the figures  because $c_B$ has only a modulating role, and this  choice does not preclude or fine-tune any particular decay channel.

Fig.~\ref{fig:constraints} illustrates the allowed values  of $|c_W|$ (left panel) and $|c_{a\Phi}|$ (right panel) in the  one-coupling-at-a-time analysis, as a function of the ALP mass and for    $f_a=1$~TeV.   Constraints from $\mathcal{B}(K\to\pi \mu\mu)$~\cite{Tanabashi:2018oca} are not displayed, since they are superseded by NA48/2 constraints on long-lived particles in $K^+\to \pi^{+} \, a(\to\mu\mu)$ decays. The grey dashed lines are projections for the SHiP experiment~\cite{Alekhin:2015byh}. 
   The figure reflects the stringent constraints from LHCb searches for displaced vertices  in the dimuon channel~\cite{Aaij:2015tna,Aaij:2016qsm,Dobrich:2018jyi} for the large mass range $m_a \in (2\mu, m_B-m_K)$,  see point 1.(a) above.   These limits are more constraining than the analogous searches performed in the kaon  sector~\cite{CERNNA48/2:2016tdo}. Remarkably, this is in contrast to the invisible scenario discussed in Sec.~\ref{sec:invisible}, for which kaon constraints are considerably stronger than those derived from $B$-meson decays  if $K^+ \to \pi^+ a$ is kinematically allowed.  These results, which take into account only one coupling at a time, could be of special interest in  specific new physics scenarios. For instance, the case of a non-vanishing  $c_{a\Phi}$ with  $c_{B}$ and $c_{W}$ disregarded (right panel) is motivated by perturbative models producing  $c_{a\Phi}$ at  tree level but $\{c_{B}, c_{W}\}$ only at loop level (e.g. $c_{B} \sim c_{W} \simeq g^2/(16 \pi^2)\,c_{a\Phi}$). Nevertheless, in all generality and for a rigorous approach, the simultaneous presence of all couplings in the electroweak bosonic basis  in Eq.~(\ref{operators}) must be considered. This may essentially modify the bounds inferred, as discussed above and illustrated next. 

Fig.~\ref{todovisible} depicts the bounds resulting when  $c_{a\Phi}$, $c_W$ and $c_B$ are simultaneously considered. Once again, in the ALP mass region in which  $B$-physics data on displaced vertices apply, they are seen to be more constraining than the bounds inferred from the kaon sector, see Figs.~\ref{todovisible}c and \ref{todovisible}d. Furthermore, 
 the four panels in the figure clearly 
 illustrate  -- for two values of $m_a$ and $c_W=c_B$ -- the remarkable pattern of constructive/destructive interference expected from the analysis in Sec.~\ref{sec:visible} and Table~\ref{tab:DestructiveVSconst}.  For instance,  the two flat directions in the photonic channel  in Fig.~\ref{todovisible}a 
  result  from destructive interference  in  both production and decay for positive $c_W/c_{a\Phi}$ and  $m_a< m_\pi$. The rest of the figures can be analogously understood. Once again, the various flat directions in different channels call for complementarity with collider data and other experimental projects. In particular, the degeneracy in parameter space which induces the flat direction in Fig.~\ref{todovisible}a and Fig.~\ref{todovisible}c, common to all rare decay channels discussed in this work, could be resolved by LHC data. Some of the flat directions appearing in ALP decays (cf.~e.g.~Fig.~\ref{todovisible}d) could also be probed by proposed beam-dump experiments such as SHiP, since they can measure ALP decays into both photons and muons, and because $\mathcal{B}(a\to \mu\mu)$ and $\mathcal{B}(a\to \gamma\gamma)$ do not simultaneously vanish. This is also true for various LHC searches, and so both experiments could be good handles on removing flat directions, though a full analysis is beyond the scope of this work.

\section{Conclusions}
\label{sec:conclusion}

The field of axions and ALPs is blooming, with an escalation of efforts both in theory and experiment.  Theoretically, the fact that no new physics has shown up yet at colliders or elsewhere  positions the SM fine-tuning issues  as the most pressing ones and 
leads to further implications for our perspective of dark matter. The silence of data is calling for a rerouting guided by fundamental issues such as the strong CP problem and an open-minded approach to hunt for the generic tell-tale of global hidden symmetries: derivative couplings as given by axions (light or heavy) and ALPs.  Experimentally,  the  worldwide program to hunt specifically for axions and ALPs is growing fast.  At the same time, other experimental programs are realizing their potential to tackle the axion and ALP parameter space, e.g.  the LHC and beam dump experiments.

In the absence of data supporting any concrete  model of physics beyond the SM, effective Lagrangians provide a model-independent tool based on the SM gauge symmetries.  Very often the effective analyses rely on considering one effective coupling at a time, though, instead of the complete basis of independent couplings. The time is ripe for further steps in the direction of a multi-parameter analysis of the ALP effective field theory, and  this is the path taken by this work. 

We have considered the impact on FCNC processes of the complete basis of bosonic electroweak ALP effective operators at leading order (dimension $5$), taking into account the simultaneous action of those couplings. As  this basis is flavor-blind, its impact on flavor-changing transitions (e.g. $d_i\to d_j a$, with $i\neq j$) starts at  loop level.   Indeed, the experimental accuracy achieved on rare-decay physics, as well as on  limits of ALP couplings to  photons, electrons, and
nucleons, is so stringent  that loop-induced contributions may provide the best bounds in a large fraction of the parameter space. 
 
We first revisited previous results in the literature, which had been derived considering just one operator at a time.  We studied next the simultaneous action of the various electroweak couplings. An interesting pattern of constructive/destructive interference has been uncovered, which depends on the relative sign of the couplings and on the channel and mass range considered. In this way, the previous very stringent bounds stemming from kaon and B-decay data are alleviated. Furthermore, LHC searches for light pseudoscalar particles have been highlighted as more important in regions where deconstructive interference weakens flavor bounds. While they are generally considerably less sensitive than flavor observables,  LHC searches are shown to provide complementary information to low-energy probes, exploring otherwise inaccessible directions in the ALP parameter space. We have also explicitly illustrated how they can overcome some of the blind directions on  rare meson decays identified here.  

We have derived the most up-to-date constraints on the effective electroweak ALP parameter space for two well-motivated scenarios: (i) an ALP decaying into channels invisible at the detector; (ii) an ALP decaying into $\gamma\gamma$, $ee$ and/or $\mu\mu$. The conclusion is that searches for $K\to \pi \nu \bar{\nu}$ decays provide the most stringent constraints in the first case. In contrast, for the second scenario, the strongest constraints  arise from searches at LHCb for long-lived (pseudo)scalars (displaced vertices) in the decays $B\to K^{(\ast)} a(\to \mu\mu)$. This  illustrates beautifully the potential of flavor-physics observables to constrain new physics scenarios. These searches will be improved in the years to come thanks to the experimental effort at NA62, KOTO, LHCb and Belle-II, providing tantalizing oportunites to discover new physics,  complementary  to the direct searches performed at the LHC.  
 
 Much remains to be done to fully encompass the ALP parameter space. For instance, the anomalous  ALP gluonic coupling has not been considered in this work. Even if it cannot mediate FCNC processes, it may impact  our results for the visible ALP via the quantitative modification of the branching ratios. In fact, recent ALP analyses of FCNC decays~\cite{Alonso-Alvarez:2018irt} take into account  the simultaneous presence of the gluonic coupling and just one electroweak ALP coupling, but no work considers all ALP bosonic couplings together, let alone the complete basis of operators including the most general fermionic ones.  This effort is very involved and will be the object of future work. In a different realm,  note that the type of effective operators considered above assumes a linear realization of electroweak symmetry breaking; the alternative of analyzing ALP FCNC processes via the non-linear effective SM Lagrangian is pertinent and also left for future consideration.

\begin{acknowledgments}
We acknowledge M.~Borsato, F.~Ertas, F.~Kahlhoefer, J.M.~No, V.~Sanz, Z.~Ligeti and M.~Papucci for very useful exchanges. O.~S. thanks the IFT at the Universidad Autónoma de Madrid for the kind hospitality.  M.~B.~G, R.~dR. and P.~Q. thank  Berkeley LBNL, where part of this work has been developed. This work has been supported by the European Union's Horizon 2020 research and innovation programme under the Marie Sklodowska-Curie grant agreements 690575 (RISE InvisiblePlus) and N$^\circ$~674896 (ITN Elusives). M.~B.~G and P.~Q.~also acknowledge support from the the Spanish Research Agency (Agencia Estatal de Investigación)
through the grant IFT Centro de Excelencia Severo Ochoa SEV-2016-0597, as well as from the ``Spanish Agencia Estatal de Investigación" (AEI) and the EU ``Fondo Europeo de Desarrollo Regional" (FEDER) through the project FPA2016-78645-P. The work of P.~Q.~was supported through a a La Caixa-Severo Ochoa predoctoral grant of Fundaci\'{o}n La Caixa.
\end{acknowledgments}
\appendix
\section{Loop factors}
\label{App:loopFactors}
The loop contributions to the  ALP decay  into photons and fermions have been computed in Ref.~\cite{Bauer:2017ris}. The loop functions in Eq.~(\ref{Eq:cgammaEff}) 
read
\begin{align}
B_0= \bigg(\sum_{{f\,=\,u,c,t}}N_c Q_f^2\,B_1(\tau_f)-\sum_{f\,=\,d,c,b,\ell^{-}_{\alpha}}N_c Q_f^2\,B_1(\tau_f)\bigg)
\end{align}
where 
\begin{equation}
   \begin{array}{l}
    B_1(\tau) = 1 - \tau\,f^2(\tau) \,, \\
    B_2(\tau) = 1 - (\tau-1)\,f^2(\tau) \,, 
   \end{array}
\end{equation}
with 
\begin{align}
f(\tau) = \left\{ \begin{array}{ll} 
    \arcsin\frac{1}{\sqrt{\tau}} \,; &~ \tau\ge 1 \,, \\
    \frac{\pi}{2} + \frac{i}{2} \ln\frac{1+\sqrt{1-\tau}}{1-\sqrt{1-\tau}} \,; &~ \tau<1 \,.
   \end{array} \right.
\end{align}
where $\tau_f\equiv 4m_f^2/m_a^2$, $Q_f$ denotes the electric charge of the fermion $f$ and $N_c^f$ is the color multiplicity ($3$ for quarks and $1$ for leptons.

\end{document}